\documentclass[a4paper,usenatbib,useAMS]{mnras}

\usepackage{graphicx}	
\usepackage{amsmath}	
\usepackage{amssymb}	
\usepackage{multicol}        
\usepackage{bm}		
\usepackage{pdflscape}	
\usepackage{color}

\usepackage[T1]{fontenc}
\usepackage{ae,aecompl}

\usepackage{txfonts}

\title[]{Diffusive Shock Acceleration and Turbulent Reconnection}

\author[C. Garrel et al.]{
Christian  Garrel$^{1,2}$,
Loukas Vlahos$^2$,
Heinz Isliker$^2$  and Theophilos Pisokas$^2$ \\
$^1$Master of Astrophysics,  Universit\'e C\^ote d\'\ Azur (MAUCA), D\'epartement de Physique de l\'\ Universit\'e de Nice \\Sophia-Antipolis and Observatoire de la C\^ote d\'\ Azur, Parc Valrose, 06100 Nice, France\\
$^2$Department of Physics, Aristotle University, 54124 Thessaloniki, Greece
}

\date{Accepted XXX. Received YYY; in original form ZZZ}

\pubyear{2018}

\begin{document}

\maketitle

\begin{abstract}
Diffusive Shock Acceleration (DSA) cannot efficiently accelerate particles without the presence of self-consistently generated or pre-existing  strong turbulence  ($ \delta B/B \sim 1 $)  in the vicinity of the shock. The
problem we address
in this article is: if large amplitude magnetic disturbances are present upstream and downstream of a shock   then Turbulent Reconnection (TR) will set  in and will participate not only in the elastic scattering of particles but also in their heating and acceleration. We demonstrate that large amplitude magnetic disturbances and Unstable Current Sheets (UCS), spontaneously formed in the strong turbulence in the vicinity  of a shock, can accelerate particles as efficiently as DSA in {\bf large scale systems and on long time scales}. We start our analysis  with ``elastic'' scatterers upstream and downstream and estimate the energy distribution of particles escaping from the shock, recovering the well known results from the DSA theory. Next we analyze the additional interaction of the particles with active scatterers (magnetic disturbances and UCS) upstream and downstream of the shock.  We show that the asymptotic energy distribution of the particles accelerated by DSA/TR has very similar characteristics with the one due to DSA alone, but the synergy of DSA with TR is much  more efficient: The acceleration time is an order of magnitude  shorter and the maximum energy reached two orders of magnitude higher.  We claim that DSA is the dominant acceleration mechanism in a short period before TR is established, and then  strong turbulence will dominate the heating and acceleration of the particles. In other words, the shock serves as the mechanism to set up a strongly turbulent  environment, in which the acceleration mechanism will ultimately be the synergy of DSA and TR. 

\end{abstract}

\begin{keywords}
particle acceleration and heating -- turbulence -- diffusive shock acceleration -- reconnection --- turbulent reconnection
\end{keywords}



\section{Introduction}
The acceleration of charged particles in space and astrophysical plasmas remains an open problem. In space plasmas the major breakthrough in understanding particle acceleration was made in the beginning of the 50s  by Fermi \citep{Fermi49, Fermi54}. Fermi proposed two  acceleration mechanisms for astrophysical plasmas. One was based on the {\bf stochastic} interaction of particles with large amplitude magnetic irregularities (``magnetic clouds'') and the second one on the {\bf systematic or regular} acceleration of particles by converging magnetic traps.

The studies following the initial   ideas proposed by Fermi gradually departed from the concepts put forward by Fermi.  The stochastic acceleration (second order Fermi acceleration) was modeled in the form of  resonant or non resonant interaction of particle with a spectrum of low amplitude $(\delta B/B<<1)$  linear MHD waves   \citep{Kulsrud71}. For more details  see the analyses presented in the reviews \cite{Melrose94, Melrose2009, Miller90, Miller97, Petrosian12}  and the references therein. 
The systematic acceleration (first order Fermi acceleration) was modeled as Diffusive Shock acceleration (DSA) \citep{Krymskii77, Axford78, Bell78a, Blandford78}. It is worth discussing briefly the weaknesses of the models used to implement the ideas proposed by Fermi. 

 The stochastic acceleration or  stochastic ``turbulent" acceleration (STA), as it is called, was modeled as the diffusion of particle energy  within a spectrum of low amplitude waves, by using the Fokker Planck  equation. The transport coefficients were estimated through the  quasilinear approximation \citep{Achterberg81}. For the STA to be efficient two conditions should be satisfied: (1) The energy of the waves should be sufficiently large and (2) the particles should have a sufficiently large velocity to resonate with the waves \citep{Melrose94}. The stochastic interaction of particles with low amplitude waves made this mechanism inefficient for the acceleration of high energy particles. The strong dependence of the index of the accelerated particles on the spectrum of the waves suggested that there is no universal index for the accelerated particles, in contrast to what is usually observed. The STA is not necessarily the correct model for the stochastic acceleration of particles by {\bf large amplitude} magnetic disturbances, as it has been shown by several authors (see \cite{Parker58, Ramaty79,   Pisokas17}). We believe that the stochastic scattering of electrons and ions off large amplitude magnetic fluctuation should be based again on the original ideas put forward by \cite{Fermi49}.
 
 The evolution of large amplitude MHD waves, needed to scatter particles in the vicinity of a shock, is a very interesting problem and has been analyzed in several articles  \citep{Biskamp89, Dmitruk03, Dmitruk04, Arzner04, Arzner06, Servidio10, Servidio11, Isliker17a} and reviews \citep{Matthaeus11, Karibabadi2013c, Karimabadi13a}. All the above studies agree that strong turbulence leads to the spontaneous formation of an environment that we will call here Turbulent Reconnection (TR), and which is dominated by large amplitude magnetic fluctuations and Unstable Current Sheets (UCS) \cite{Matthaeus86, Lazarian99, Lazarian15}. The interaction of electrons and ions with plasmas in the TR state has been analyzed by \cite{Ambrosiano88, Dmitruk04, Arzner06,   Vlahos16,Pisokas17,Isliker17, Pisokas18}.

 The acceleration of particles in the vicinity of a shock is a prominent acceleration mechanism for astrophysical plasmas. The details on how this mechanism will operate in different cosmic environments remain an open and very complex problem (see the reviews \cite{Drury83, Burgess12, Schure12}). 
 
 In the heliosphere, the Coronal Mass Ejection (CME) shock, the Earth's and planetary Bow Shocks interact with the Solar Wind, which is already in a TR state \citep{ Zank14, Matthaeus15, Chasapis15, Osman14, Khabarova16, Khabarova17}. Downstream of the heliospheric shocks the amplitude of the magnetic fluctuations becomes even stronger and the formation of a TR environment is easier to establish. The presence of a TR environment in the Heliospheric Termination Shock is under intense discussion \cite{Lazarian09, Drake10, Burgess16}.

 The role of Shock acceleration during Supper Nova Remnants (SNR) and its importance for accelerating cosmic rays is still an open problem in High Energy Astrophysics. A major concern in the theory of particle acceleration through DSA in SNR is the mechanism which traps the accelerated particles in the vicinity of the shock surface. The large amplitude magnetic fluctuations upstream and downstream of the shock can be driven by different MHD or kinetic instabilities (see the review of \cite{Bykov13}). The interaction of the shock precursor with  density fluctuations in the pre-shock media \citep{delVale16} or the post-shock turbulence arising from various instabilities \citep{Balsara01, Balsara05, Giacalone07} can create large amplitude MHD perturbations. \cite{Bell04} proposed that turbulence upstream can be excited by current driven instabilities of accelerated particles. 
 The most advanced recent simulations (\cite{Caprioli14a, Caprioli14b, Caprioli14c, Bai15, vanMarle18}) so far cannot capture the large spatial scales and the long term evolution of the excited instabilities and interactions involved in realistic astrophysical systems, i.e.\ their results cover only the microphysics of the very early moments of the shock acceleration.

 The link between shock and TR has been  analyzed by \cite{Karimabadi2014}, using large scale PIC simulations.  The presence of   reconnecting current sheets in the vicinity of the shock and their role in the acceleration of particles has been analyzed also in depth by several authors (see \cite{Zank15, Matsumoto15, leroux16}). The importance of turbulent reconnection in the vicinity of the shock and its role in heating and accelerating particles is currently a very important problem waiting its solution.  
 
 In this article we start by analyzing DSA, assuming scatterers upstream and downstream that are elastic. We then move on to study the role of stochastic acceleration in DSA, by assuming that the scatters are active, mostly in the form of large amplitude magnetic fluctuations. In the final step we assume that a percentage of the scatterers are UCS, modeling a TR state of the plasma in the vicinity of the shock. In the current literature the case of DSA in the presence of UCS has been studied \citep{Zank15, leroux16}, as well as 
 the acceleration of particles by TR in the absence of a shock \citep{Pisokas18}.

 The structure of our article is as follows: in section 2 we review briefly the basic theoretical results of DSA, and in section 3 we propose a large scale model for DSA. In section 4.1 we present our results  for passive scatterers, and in Sect.\ 4.2 and 4.3 for active scatterers,  assuming that their interaction is either stochastic or stochastic and systematic.

\section{Diffusive Shock Acceleration}

\label{Section:DSA}

Diffusive Shock Acceleration is based on the assumption that particles can be confined near the shock discontinuity by scattering off Magnetohydrodyanmic large scale 
disturbances that act as scattering centers. The scattering mean free path $\lambda_{sc}$ is assumed to be much larger than the shock thickness and much shorter than the length $L$ of the area where the wave activity upstream and downstream is strong. As a result particles can cross the shock discontinuity repeatedly. The charged particles gain energy through the repeated scatterings off the converging up- and downstream scattering centres. 

We define the shock compression ratio as 
\begin{equation} \label{comp}
	r=\frac{\rho_1}{\rho_2}=\frac{U_1}{U_2}
\end{equation}
where $\rho_1, U_1$ are the density and the fluid velocity upstream of the shock, and $\rho_2,U_2$ the ones downstream.
The energy spectrum of the particles accelerated by the shock reaches asymptotically the distribution \citep{Drury83}
\begin{equation} \label{Distr}
	f(p) \sim p^{-3 U_1/(U_1-U_2)} \sim p^{-3r/(r-1)} ,
\end{equation}
where $p$ is the momentum of the particles. For strong shocks the compression ratio is $r \sim 4$ and 
$3 r/(r-1) \sim 4.$ The fact that the exponent depends only on the compression ratio and reaches asymptotically the value of 4, explains the attractiveness of DSA for astrophysical plasmas.

The analysis of the energy gain $\langle \Delta W \rangle$ of test particles crossing the shock discontinuity yields (see \cite{Longair11})
\begin{equation}
\label{eq:dEm}
\frac{\langle   \Delta W \rangle}{W}  \sim  \frac{Uu}{2c^2}
\end{equation}
where $U$ is the shock velocity, $u$ the velocity of the particle,
and $W$ the total energy of the particle. 
From the downstream region the particle can recross the shock, gaining another increment in energy $(Uu/2c^2)$. The total energy increase for a particle making one round trip is therefore on average
\begin{equation}
\label{eq:dEm2}
\frac{\langle  \Delta W \rangle  }{W}  \sim \frac{Uu}{c^2}.
\end{equation}
In the downstream region, the scattering process insures that the particle distribution is isotropic, but occasionally, the flow can sweep particles away from the shock. Particles can thus escape from the downstream area with a finite probability $P_{esc}$. This probability can be calculated as (see \cite{Achterberg08})
\begin{equation}
\label{eq:pesc}
P_{esc} \approx \frac{U}{c}.
\end{equation}
We can now introduce $t_{up}$, $t_{dwn}$, and $t_{cycle}$, respectively, as the times a particle spends on average in the upstream and the downstream region, and the time of a cycle of one round trip across the shock \cite{Drury83}
\begin{align}
\label{eq:tup}
t_{up}  \sim \frac{\lambda_{sc_{up}}}{U}\\
\label{eq:tdwn}
t_{dwn} \sim \frac{\lambda_{sc_{dwn}}}{U/4}\\
\label{eq:tcylc}
t_{cycle} = t_{up} + t_{dwn} \sim  \left(\frac{\lambda_{sc_{up}}}{U} + \frac{\lambda_{sc_{dwn}}}{U/4}\right) \sim \frac{\lambda_{sc_{up}}}{U}
\end{align}
(ignoring factors of 2).
Assuming that the particles execute a random walk upstream and downstream along their path between the scatterers, the mean square displacement is given as 
\begin{equation} \label{rwk}
	<R^2>=6D_{up/dwn} t
\end{equation}
where $D_{up/dwn}$ is the diffusion coefficient upstream and downstream, respectively. We can estimate the relation of the mean free path to the diffusion coefficient for relativistic particles if we assume that $t \sim \lambda_{sc_{up/dwn}}/c$ 
\begin{equation}
	\label{lambda_D}
	\lambda_{sc_{up}}  \sim \frac{D_{up}}{c},
	\ \ \ 
	\lambda_{sc_{dwn}} \sim \frac{D_{dwn}}{c}
\end{equation}

\begin{figure*}
(a)
 \includegraphics[width=\columnwidth]{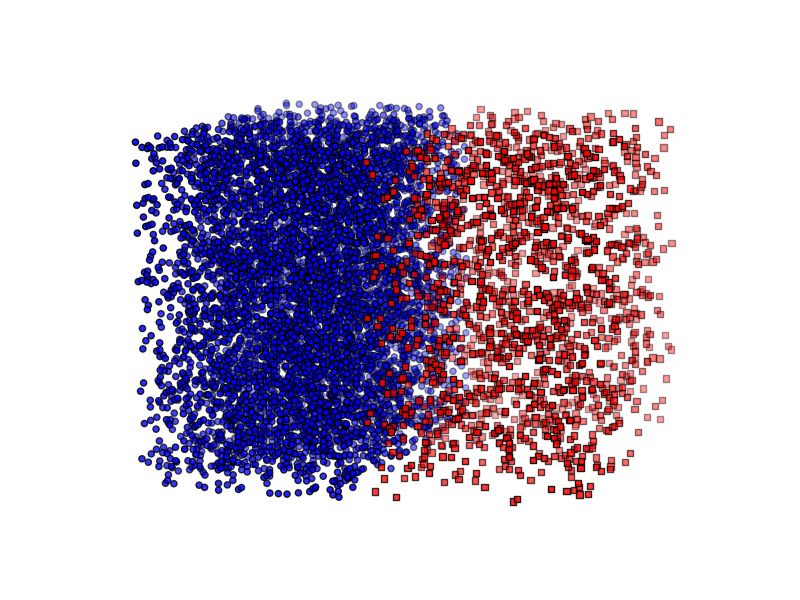}
(b)
 \includegraphics[width=\columnwidth]{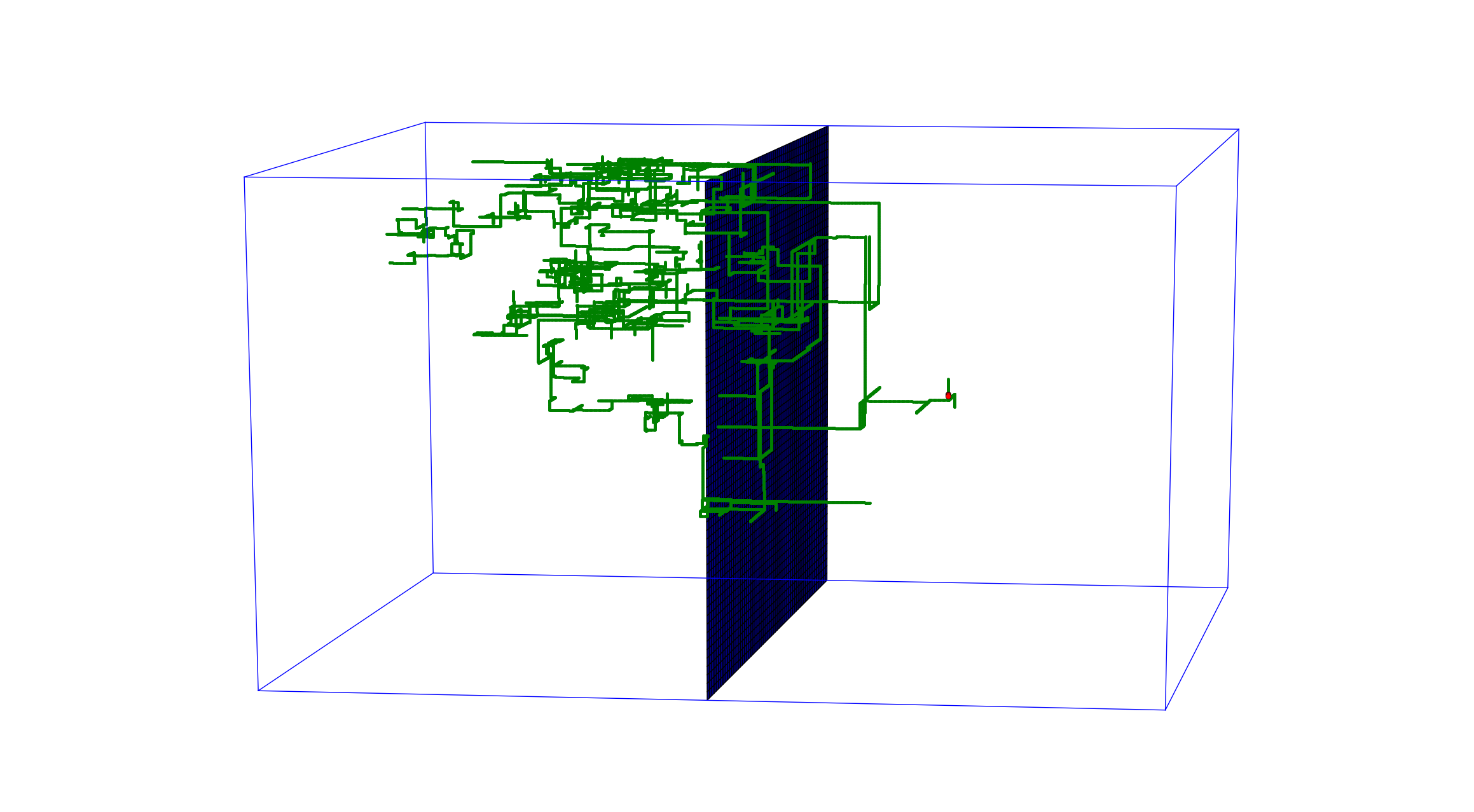}
 (c)
 \includegraphics[width=\columnwidth]{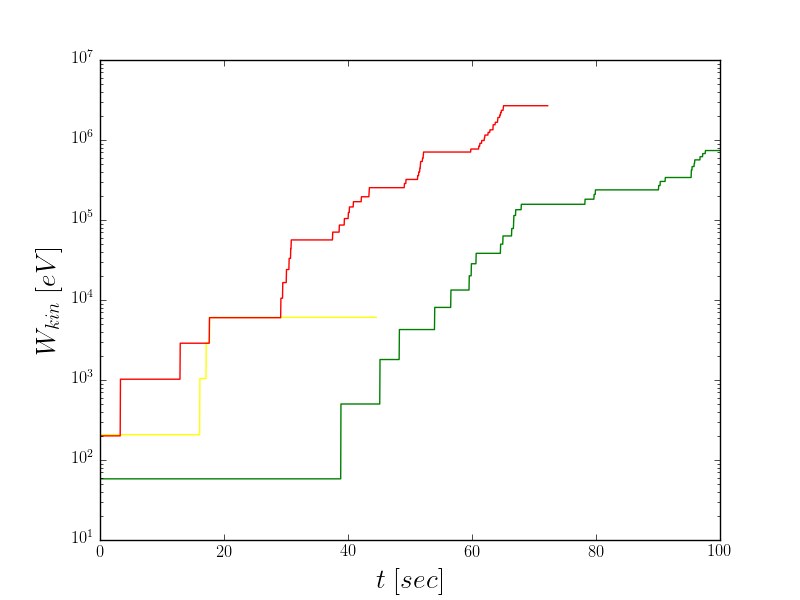} 
 (d)
  \includegraphics[width=\columnwidth]{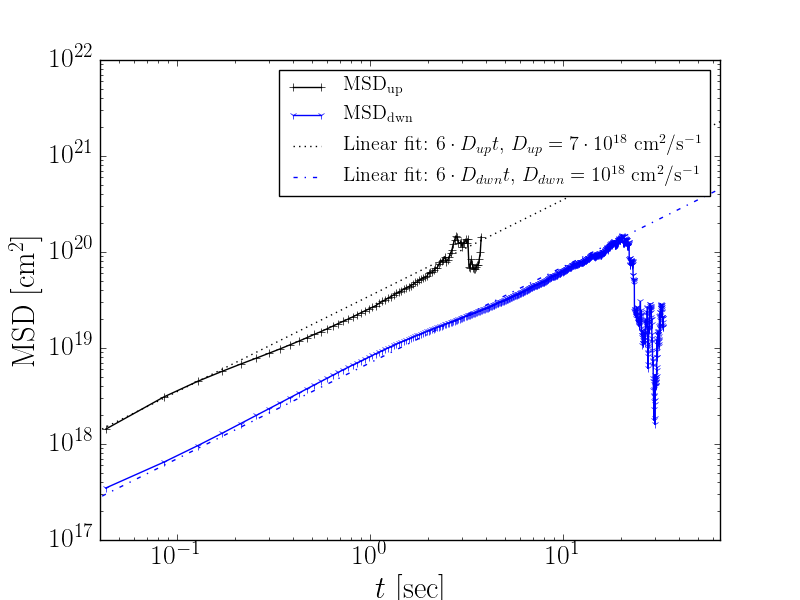} 
 \caption{(a) Small version of the simulation box with the planar shock wave in the middle.(b) Trajectory of a typical electron inside the simulation box. (c) The kinetic energy increase of typical electrons traveling inside the simulation box as a function of time. Electrons gain systematically energy as they cross the shock.
 (d) Mean square displacement of the electrons upstream (black) and downstream (blue).}
 \label{fig:1}
\end{figure*}

We notice that we can also re-write $t_{cycle}$ as 
\begin{equation}
\label{eq:tcycle2}
t_{cycle} \approx \frac{4}{c} \left(\frac{D_{up}}{U_{1}} + \frac{D_{dwn}}{U_{2}}\right) \sim \frac{D_{up}}{cU}
\end{equation}
for relativistic particles \citep{Drury83}, with $U_1$ and $U_2$ being defined in Eq.\ (\ref{comp}).

The rate of the energy gain of the particles is 
\begin{equation} \label{RateEnergy}
	\frac{\langle   \Delta W \rangle }{dt} \sim \frac{U^2}{D_{up}}W\sim \frac{W}{t_{acc}}
\end{equation}
(from Eq.\ (\ref{eq:dEm2}), with $dt = t_{cycle}$ from Eq.\ (\ref{eq:tcycle2}), and by assuming relativistic particle velocities),
which defines the acceleration time $t_{acc} \sim D_{up}/U^2$.

Finally, the energy distribution has been estimated to be
\begin{equation}
	\label{eq:imp}
	N(E) \propto  E^{-(1+ t_{acc}/t_{esc})},
\end{equation}
(see \cite{Jones94, Achterberg08} ), with the escape time $t_{esc} = t_{cycle}/P_{esc} \sim [D_{up}/cU]/[U/c] \sim D_{up}/U^2 \sim t_{acc}$ (see Eq.\ (\ref{eq:pesc}), and by using $t_{cycle}$ as a characteristic time scale) and $t_{acc}$ from Eq.\ (\ref{RateEnergy}).  

\section{The Numerical Model}

\label{num_model}

We construct a 3D grid ($N \times N \times N$) with linear size $L$. The box is separated into two parts that we call downstream and upstream region, and which represent, respectively, the area behind and in front of the shock discontinuity that itself is located at the middle plane. The grid is randomly filled by $N_{sc_{dwn}}$ scatterers downstream and $N_{sc_{up}}=N_{sc_{dwn}}/r$ scatterers upstream, where $r$ is the compression ratio (see Eq. \ref{comp}). We define $R_{up/dwn} = N_{sc_{up/dwn}}/[N^2 \times (N-1)/2]$ as the fraction of grid point which are scatterers.  The mean free path the particles travel between two scatterers is different on the two sides of the discontinuity 
(see Fig.~\ref{fig:1}(b)) and it can be determined as $\lambda_{sc_{up/dwn}} = \ell/R_{up/dwn}$, with $\ell= L/(N-1)$ the grid width. The time between two shock crossings is $\Delta t=t_{cycl}=t_{up}+t_{dwn}$ with $t_{up/dwn}$ the mean time particles stay in the upstream/downstream region. At time $t=0$ the injected distribution $n(W,t=0)$ is a Maxwellian with temperature $T$.

The scatterers can be either elastic or active. In the case the interaction of the particles with the scatterers is elastic, the scatterers only affect the direction particles move after their interaction with the scatterer. When the scatterers are active a particle loses or gains energy in its interaction with the scatterers. Energization of the particles occurs in any case when the particles are crossing the shock wave. In 
Fig.~\ref{fig:1}(a) we show a small version (for better visualization) of the box with the downstream scatterers in blue and the upstream scatterers in red. The two upstream and downstream boundary planes of the box that are parallel to the shock surface are always open, and the four boundary planes that are perpendicular to the shock surface can be periodic (partially open box) or open (open box). The particles have a higher probability to escape from the downstream region than from the upstream region. We choose the initial position of the particles randomly and only in the upstream region, and everywhere the particles are bound to follow the grid-lines in their motion. The direction particles move when they are upstream is randomly chosen only between 5 directions: the four directions parallel to the shock and the opposite direction to the shock normal, i.e.\ towards the shock front. A particle can escape from the upstream region e.g.\ when it moves from downstream to upstream and never meets a scatterer upstream, yet the probability for this to happen is very small. On the other hand, when particles are downstream, all directions of motion are equally probable when a particle encounters a scatterer. A typical trajectory of a particle inside the simulation box is shown in Fig.~\ref{fig:1}(b). 

The parameters used in our simulations are related to the plasma parameters in the  upper corona. We choose the magnetic field to be $B=10$ G, the density of the plasma $n_{0}=10^{7}$ $cm^{-3}$, the ambient temperature around $100$ eV, the length $L$ of the simulation box is $10^{10}$ cm, $R_{dwn}=10\%$, and we use $N=601$ grid points in each direction. The Alfv\'en speed is comparable with the thermal speed of electrons ($V_{A} \approx 7 \cdot 10^{8}\; cm s^{-1}$). The mean free path of the particles upstream is $\lambda_{sc_{up}} \; = L /[(N-1) R_ {up}]  \sim   10^9 cm.$  The shock wave velocity is chosen as $U=M_A \cdot V_{A}$ with the Alfv{\'e}n Mach number $M_A=5$. Each shock crossing changes the energy of the particles according to Eq.\ (\ref{eq:dEm}). Then, for relativistic particles ($\geq 1$ MeV), the typical energy increment is of the order of $(\Delta W/W) \approx U/c\sim 10^{-1}$ for one round trip across the shock.

\section{Results}

\subsection{Elastic scattering}
\label{sect:elscat}

In this section we assume that the scatterers affect only the direction along which the particles travel inside the simulation box, and the box is partially open, only the upstream and downstream boundary of the box are open. The temporal evolution of the kinetic energy of typical electrons is presented in Fig. ~\ref{fig:1}(c). The particles systematically gain energy, which though happens only when they cross the shock discontinuity. 

Our aim here is to compare our numerical results with the theoretical arguments and results presented in Section \ref{Section:DSA}, so we use relativistic particles for our approximate estimates. It is important to stress that the relativistic particles appear in our simulations around 20 sec after the injection of the thermal particles.

In Fig. \ref{fig:1}(d) we show the mean square displacement of the particles 
$<R^2>$ as a function of time downstream and upstream. The diffusion of the particle is normal, as we would expect it for a random walk process, and a linear fit gives the estimates of the diffusion coefficients
$D_{up} \approx  8 \times 10^{18}cm^2 s^{-1}$ and $D_{dwn} \approx 10^{18} cm^2 s^{-1}.$  The value of 
$\lambda_{sc_{up}} \approx D_{up}/U \sim  10^9 cm$, as yielded by Eq.\ (\ref{lambda_D}),
is thus in good agreement with the value that follows from the characteristics of our simulation set-up, see Sect.\ \ref{num_model}.

\begin{figure*}
(a)
 \includegraphics[width=\columnwidth]{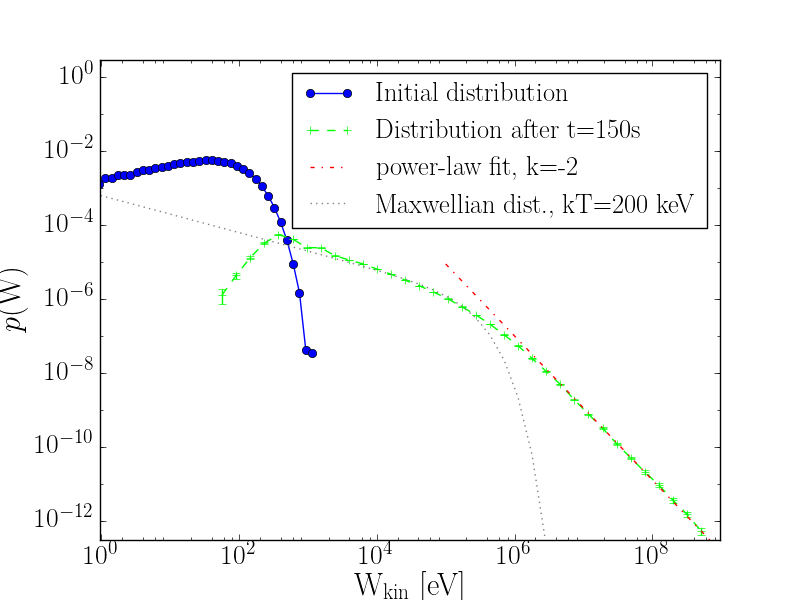}
(b)
 \includegraphics[width=\columnwidth]{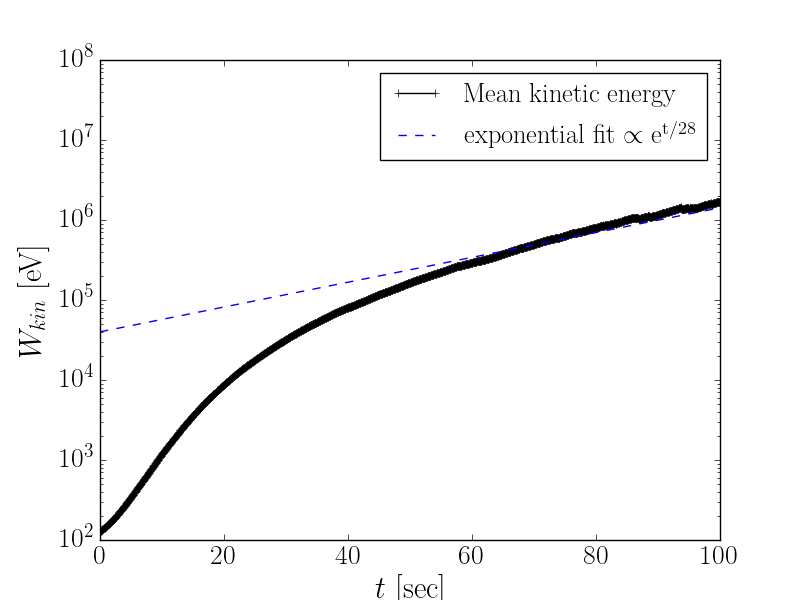}
(c)
 \includegraphics[width=\columnwidth]{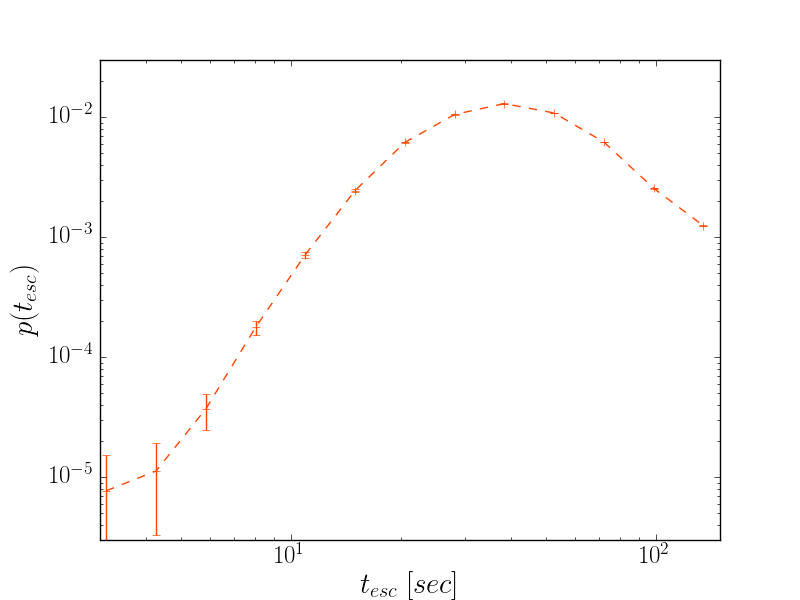}
 (d)
  \includegraphics[width=\columnwidth]{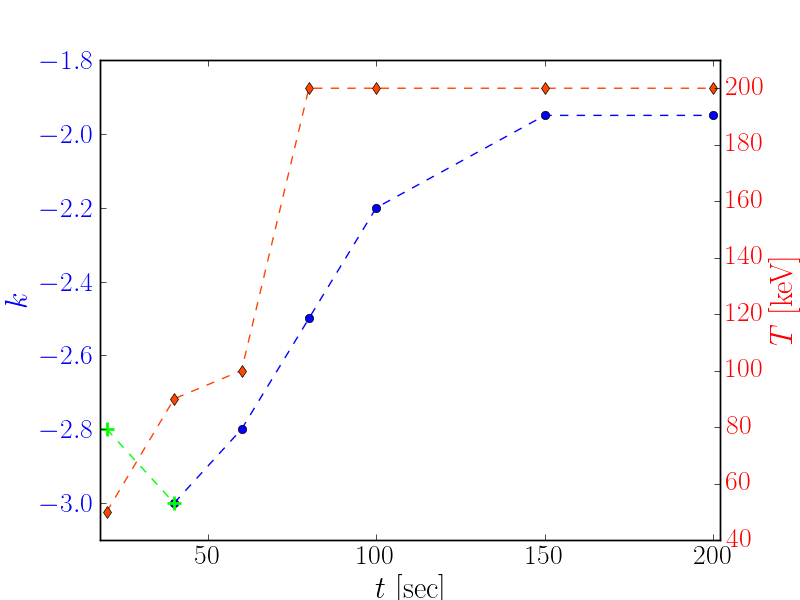} \caption{(a) The asymptotic kinetic energy  distribution of the electrons escaping from the box downstream: Initial energy distribution (blue), particles that have escaped from the box (green), together with a power-law fit $W^k$ with  $k=-2$ (red). (b) The evolution of the mean energy as a function of time (black) with an exponential fit (blue). (c) Distribution of the escape times of the electrons. (d)  Parameters of the evolution of the energy distribution of electrons escaping downstream: The evolution of the power law index of the  relativistic electrons (blue; the green color is for the power-law tail before particles have become relativistic), and the evolution of the temperature of the non-relativistic electrons (red). }
 \label{fig:2}
\end{figure*}


The average energy gain ratio for particles 
in one complete 
cycle around the shock surface is estimated from our simulation as $<\Delta W>/W \approx 0.1$, which is close to the theoretically expected value $U/c$ (see Eq.\ (\ref{eq:dEm2}) for the relativistic case).
 
From Eqs.\ (\ref{eq:tup}) and (\ref{eq:tdwn}) we can estimate the time the particles spend upstream and downstream before they cross the discontinuity,  $t_{up_{th}}, t_{dwn_{th}} \approx 0.26 sec$, and comparing this with the values estimated from our simulation, we find $t_{up_{num}} \approx 0.19 sec$ and $t_{dwn_{num}} \approx 0.39 sec$. A typical particle completes a cycle around the shock discontinuity in $t_{cycl_{num}} \approx 0.58 sec$, while when using the mean free path or the diffusion coefficients (Eqs. (\ref{eq:tcylc}) and (\ref{eq:tcycle2})) we estimate  $t_{cycle_{th}}\approx 0.46$ sec. All the values from test particle approximation are thus in close agreement with the theoretical results in Sect.\ \ref{Section:DSA}.

 We can also compare the estimate of $P_{esc}$ from our model with the one predicted by the theory (see Eq.\ \ref{eq:pesc}). We notice that  the escape probability is energy independent, so for the estimate of $P_{esc}$, it is relevant to take the median value of $t_{cycle}$ of all particles and not just the relativistic ones. In this case we find $t_{cycle}\approx 6.6sec$, and using the relation $P_{esc} = t_{cycle}/t_{esc}$ \citep{Achterberg08}  we have $P_{esc}\approx 0.15sec$, close to the theoretical value ($\approx 0.11$).

 The energy distribution of the electrons that have escaped after $150$ seconds is presented in Figure~\ref{fig:2}(a). We find that the energy distribution of the non relativistic particles can be fitted with a Maxwelian distribution, and the high energy tail with a power-law with an index $-2$, as it is well known from theoretical arguments given in the literature and reported briefly in section 2. (Concerning the different slopes of the two Maxwellian distributions in Figure~\ref{fig:2}(a) at low energies, we note that the initial velocity is generated obeying a 3D Maxwellian, the particle motion though is bound to the grid so the final distribution follows a 1D Maxwellian.) 
 
 The temporal evolution of the mean kinetic energy is shown in Fig.~\ref{fig:2}(b). As we expect from Eq.\ (\ref{RateEnergy}), the energy increases exponentially after an initial transient period. We can  estimate the acceleration time from the simulation through an exponential fit, which yields $t_{acc_{num}}\approx 28 sec$, almost an order of magnitude slower  than the estimated theoretical value $t_{acc} \sim D_{dwn}/U_2^2 \sim 2 secs$
 (see Eq.\ Eq.\ (\ref{RateEnergy})). This is probably related with the fact that all the particles, including the thermal ones, were taken into account for the numerical estimate, whereas the theoretical estimate was derived by making the assumption that all particles have relativistic energies. The mean $t_{esc}$ is estimated to be $\sim 42$ secs.


\begin{figure}
 \includegraphics[width=\columnwidth]{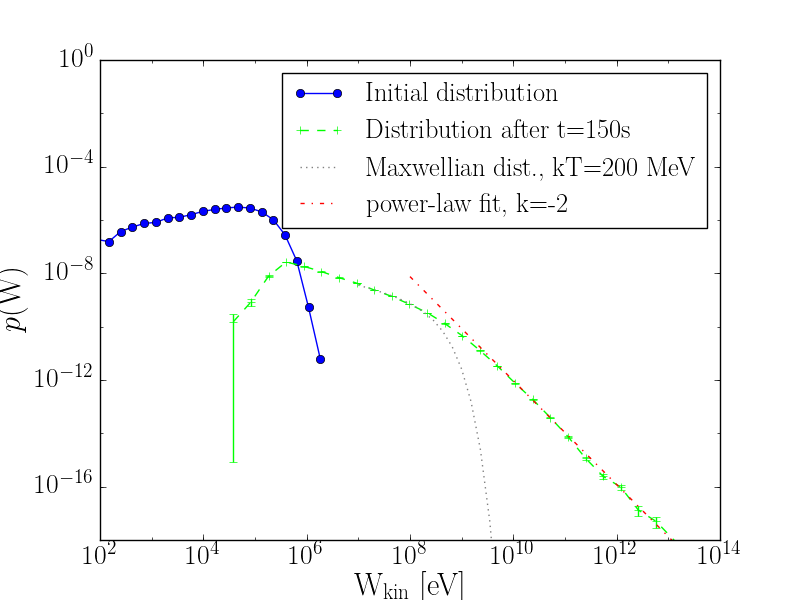}
 \caption{Energy distribution of the ions escaping from the partially open simulation box after reaching their asymptotic state ($\sim 150$ seconds) (green) and initial energy distribution (blue). The asymptotic energy distribution is a mixture of hot plasma with temperature $T= 200$ $MeV$ (gray) with a power-law distribution with index $k= -2$ (red).}
 \label{fig:ions_passive}
\end{figure}

Following the time evolution of the energy distribution of the escaping particles 
we can estimate the evolution of the temperature of the non relativistic particles and the power law index of the relativistic particles (see Fig. \ref{fig:2}(d)).  The energy distribution reaches its asymptotic shape approximately 100-150 secs after the injection of the initial Maxwellian distribution with temperature $\approx 100$ eV. The escaping distribution attains a temperature around $200$ KeV, and the high energy tail has a power law index $-2$ and extends to the maximum energy of $\sim 1$ GeV, for the electrons considered here. 
Keeping all other physical parameters fixed (plasma density, magnetic field strength, Mach number, etc), the crucial parameter that affects most of the estimates reported here is the scattering mean free path, i.e.\ the density of the large scale magnetic disturbances that scatter the particles. 
 As we can see from Eq.\ (\ref{lambda_D}), the diffusion coefficient is directly influenced by the scattering mean free path, which will affect the acceleration time. This result has been confirmed by our simulations.

Most of the theoretical estimates reported in section 2 are independent of the particles mass, so the acceleration time, the diffusion coefficients, the mean free path, etc, are similar for the ions  and the electrons 
and as estimated 
above. 
The energy distribution of the ions escaping downstream has the same characteristics with the energy distribution of the electrons (see Fig. \ref{fig:ions_passive}). The energy distribution is again a Maxwellian distribution with a power law tail with index $k\approx-2$. The difference between the energy distribution of the ions and the electrons is that the ions are hotter ($T\approx 200$ MeV) than the electrons downstream, and their power-law tail reaches higher energies ($\sim 1000$ GeV). 
\begin{figure}
 \includegraphics[width=\columnwidth]{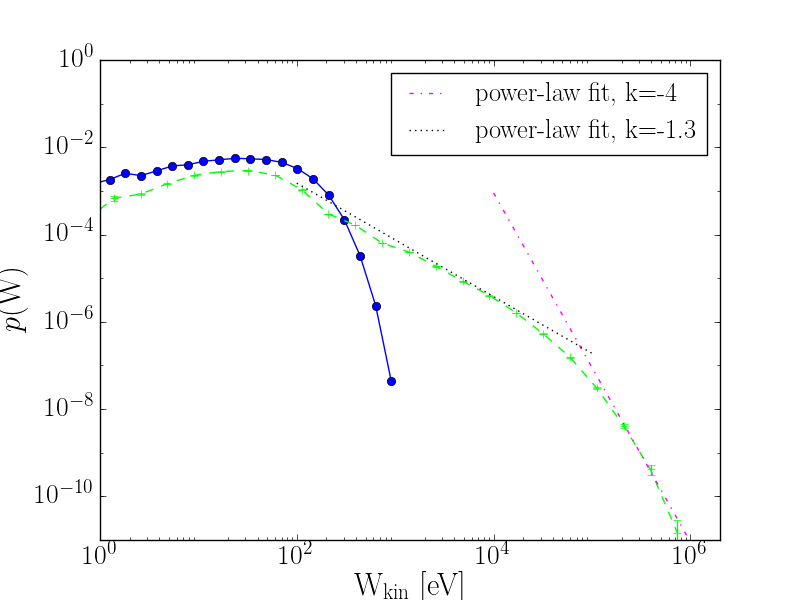}
 \caption{Initial energy distribution (blue) and the energy distribution of the electrons escaping from the box after $40$secs (green), with two power-law fits $\propto W^k$ with $k=-1.3$ (black) and $k=-4$ (red), in an open simulation box, for the case of Diffusive Shock Acceleration.}
 \label{fig:dist_open}
\end{figure}
 
 So far the simulation box used was partially open. For an open simulation box 
we again numerically estimate  the diffusion coefficients, and we find them to remain unchanged, therefore $t_{cycle}$ also stays the same ($\approx 0.6$ seconds).
The main  differences  with the partially open box are the mean escape time  and the  acceleration time. Electrons barely reach relativistic energies (particles never exceed $1$ $MeV$) and the energy distribution of particles escaping from the simulation box reaches an asymptotic state  after around $40$ seconds. The energy distribution shows moderate heating at the low energies and then forms a double power-law, with index $\approx -1.3$ at intermediate energies and index $\approx -4$ at the high energy tail (see Fig.~\ref{fig:dist_open}). This distribution is similar to the energy distribution from the partially open box reached after 40 secs, as one can see from Fig.\ \ref{fig:2}(d). In other words, the electrons  escape from  the open simulation box before attaining to the asymptotic state reached in the partially open box. 

In the case of ions in the open box, we obtain the same shape of the energy distribution as for the electrons (a Maxwellian at the low energies and a double power law at the high energies), yet the ions become hotter  ($T\approx 30$ $MeV$), the double power law tail is similar to the one of the electrons, and the  maximum energy the ions reach is $\sim 1$ $GeV$. In order with the open simulation box to get results similar to the ones from the partially open box, we would have to increase the trapping of the particles in the box by reducing the mean free path of  their interaction  with the scatterers.

Our main finding so far is that 
the accelerated electrons and ions have the same
shape of the energy distribution, namely a mixture of hot plasma with a power law distribution at the high energy tail. In order for the power law tail to assume the well known asymptotic index (-2), the particles must reach relativistic energies. 

The numerical estimates and the analysis reported here are in good agreement with the well known theoretical estimates for the DSA reported in the literature (see Sect.\ \ref{Section:DSA}).  In the next paragraph we use the same environment as above, with the only difference that the scatterers are considered to be active. We assume that large amplitude magnetic fluctuations in the vicinity of the shock interact stochastically with the electrons and ions, as it was proposed and analyzed initially by Fermi \citep{Fermi49}.

\subsection{The role of Stochastic Fermi Energization for Diffusive Shock Acceleration}

 The fact that high amplitude magnetic fluctuations are present in the vicinity of a shock implies 
 that they may also contribute substantially as scatterers 
 to the acceleration of electrons and ions. 
 These large amplitude magnetic fluctuations are not  waves, and their interaction with particles does not obey  the quasilinear wave-particle approximations. For the analysis of the interaction of electrons and ions with magnetic disturbances with $\delta B/B \geq 1$, we return to the analysis done initially by Fermi (see \cite{Parker58, Ramaty79, Miller92, Pisokas17}).

 The stochastic energy gain or loss of particles interacting with the scatterers upstream and downstream is 
\begin{equation}
\frac{\Delta W}{W} \approx \frac{2}{c^{2}} \left(U_{1/2}^{2} - \textbf{U}_{1/2} \cdot \textbf{v}\right),
\label{eq:energy}
\end{equation}
where $U_{1}$, $U_{2}$ is the velocity of the flow, respectively, upstream and downstream.
The initial distribution of the particles is again a Maxwellian with temperature $100\, eV$ (the ambient temperature of the plasma). At $t=0$, all particles are located upstream at random positions on the grid and the initial direction of their motion follows the same rules as in the DSA model with the elastic scatterers discussed above in Sect.\ \ref{sect:elscat}. As the grid constrains the motion of the particle, the term $\textbf{U}_{1/2} \cdot \textbf{v}$ has a sign randomly chosen among three possible values:  -1 for head-on scatterings, 1 for overtaking scatterings, and 0 for perpendicular scatterings. The typical energy increment is then of the order of $(\Delta W/W) \approx (U_{1/2}/c)^{2} \sim 10^{-3}$, so the energy gain is smaller per scattering off a magnetic fluctuation than in an energization event at the shock, but the number of interactions is much higher.

\begin{figure*}
(a)
 \includegraphics[width=\columnwidth]
  {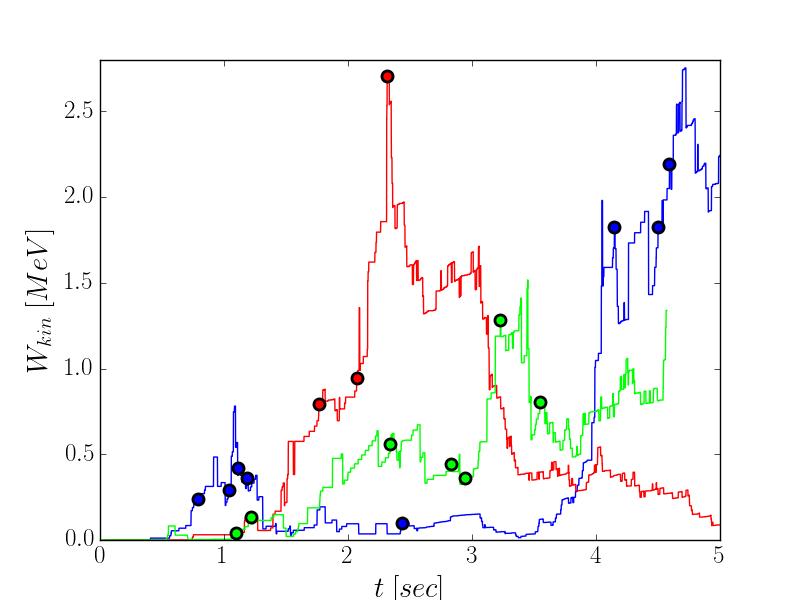}
(b)
 \includegraphics[width=\columnwidth]
 {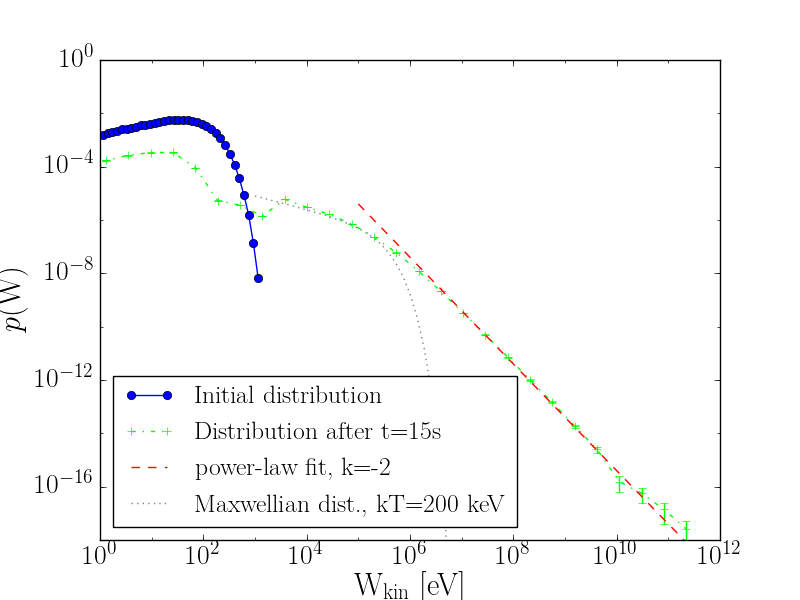}
 (c)
 \includegraphics[width=\columnwidth]{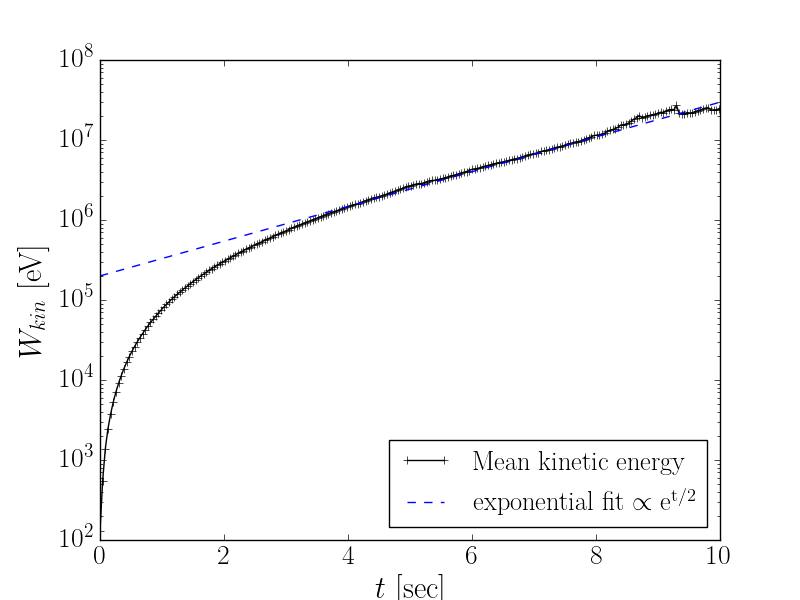}
(d)
 \includegraphics[width=\columnwidth]
 {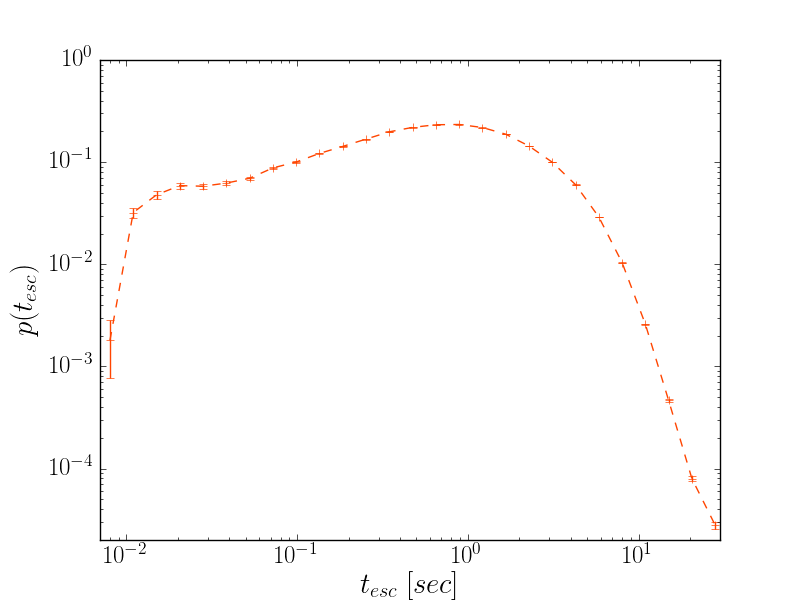}
(e)
 \includegraphics[width=\columnwidth]{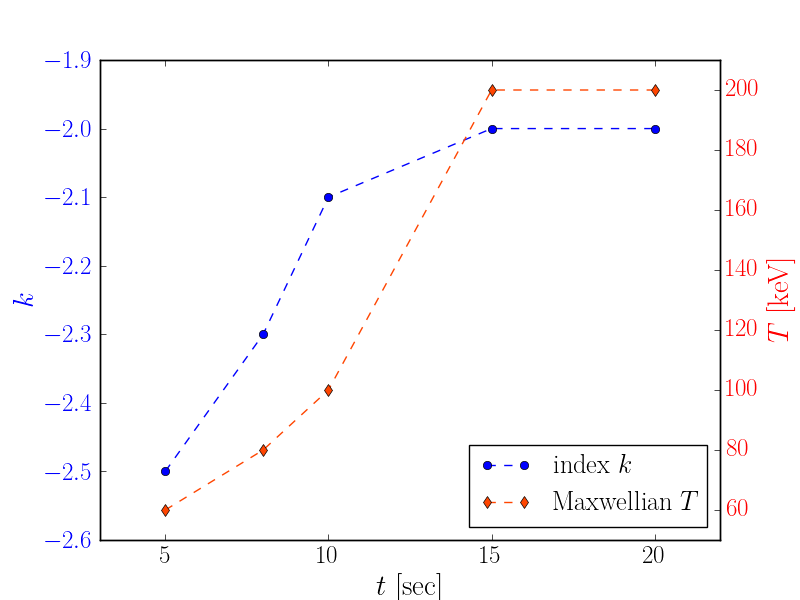}
(f)
 \includegraphics[width=\columnwidth]{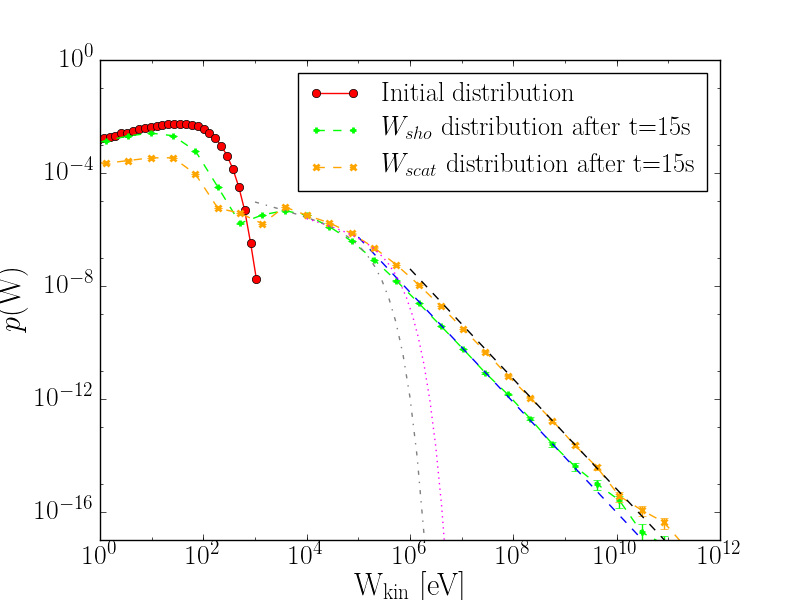}
\caption{Characteristics of Diffusive Shock Acceleration in combination  with stochastic interaction of electrons with magnetic disturbances: (a) Kinetic energy as a function of time for 3 typical particles, the circles mark shock crossings. (b) Initial kinetic energy distribution (blue) and kinetic energy distribution after 15 seconds for particles escaping from the box (green), with a power-law fit ($= W^{-2}$ (red)), and a Maxwellian distribution with temperature $T= 200$ $keV$ (gray).    (c) Mean kinetic energy as a function of time (black) with an exponential fit with slope $\approx   2 secs$  (blue). (d) Distribution of the escape times $t_{esc}$. (e) Evolution of the index $k$ of the power-law tail and of the temperature $T$ of the Maxwellian fit. (f) Initial kinetic energy distribution (red), together with the kinetic energy distribution from the shock contribution only (green) and from the scattering contribution only (orange), together with power-law fits with indexes $k= -2$ (blue and black), and Maxwellian distributions with temperature $T= 80$ $keV$ (gray) and $T= 200$ $keV$ (pink).}
 \label{fig:mix}
\end{figure*}

The kinetic energy as a function of time for three typical electrons is shown in Fig.\ \ref{fig:mix}(a). The electrons gain or lose energy stochastically when they interact with the magnetic disturbances and they systematically gain energy when they cross the shock at the times marked with circles in Fig.\ \ref{fig:mix}(a).

In the asymptotic state, the kinetic energy distribution of the electrons escaping from the open simulation box  is a mixture  of hot plasma with $T\approx 200$ $keV$, and a power law tail with index $k=-2$ (see Figure~\ref{fig:mix}(b)). We note that both $T$ and $k$ progressively increase until they reach their asymptotic value (see Figure~\ref{fig:mix}(e)). The simulation box is chosen open since in a partially open box SDA together with stochastic scatterers leads to an over-efficient energization, and we would have to lower the density of the scatterers. Since we preferred to keep the parameters as  before in Sect.\ \ref{sect:elscat}, we allow for faster escape from an open box, which reduces the energization to a reasonably efficient level.

The escape times $t_{esc}$ and acceleration times $t_{acc}$  are shown in Fig.\ \ref{fig:mix}(c) and \ref{fig:mix}(d). The acceleration time is estimated by fitting the mean kinetic energy as a function of time with an exponential (see Figure~\ref{fig:mix}(c)), which yields $t_{acc}\approx 2 secs$ that is an order of  magnitude faster than in the pure DSA reported above. We note that the median escape time ($t_{esc}\approx 2.5 secs$) is also an order of magnitude smaller than in the case of pure DSA. The synergy of stochastic acceleration by large amplitude magnetic fluctuations and of systematic acceleration at the shock influences the efficiency of the electrons' acceleration and increases the maximum energy they reach by two orders of magnitude ($\approx 100$ $GeV$). 

We keep track separately of the energy particles gain along their path from the scatterers ($W_{scat}$) and from their crossings of the shock surface 
($W_{shock}$, the classical DSA result). Indeed the distribution of $W_{scat}$ is a mixture of hot plasma with temperature $T\approx 200\,$keV, which coincides with the low energy part of the final total distribution, and a power law tail with index $k\approx -2$. The distribution of $W_{shock}$ has the same general form as the energy distribution due to the interaction of the particles with the magnetic fluctuations, but with a lower temperature $T\approx 80\,$keV (see Figure~\ref{fig:mix}(f)). Also, the ratio of the mean  integrated  energy  gain from the scatterers and the shock  $\left\langle W^t_{scat} \right\rangle / \left\langle W^t_{shock} \right\rangle$ is $\approx 6.6$, i.e.\ the contribution of $W_{scat}$ is around seven times bigger than the contribution of $W_{shock}$. So, the stochastic interaction between magnetic fluctuations and particles dominates their heating and is equally effective in  the acceleration of the high energy tail. At the same time it retains the standard DSA characteristics for the heated and the accelerated particles. In most hybrid, PIC and MHD(PIC) simulations (see for example \cite{Caprioli14a, Caprioli14b, Caprioli14c,  vanMarle18} one of the main arguments given that the simulations were successful to reproduce the results expected from DSA is the fact that the power law index of the high energy tail is close to -2. The combined effect of stochastic and systematic acceleration reported here maintains this result and influences mainly the efficiency of the accelerator.

The fact that the DSA and stochastic acceleration by large amplitude magnetic disturbances can be similar was already pointed out by \cite{Jones94}. He compared separately the two mechanisms and concluded that "DSA has no particular advantage over stochastic acceleration". In our opinion the synergy of the two mechanisms analyzed here is even more appealing and much more efficient, and, as mentioned, 
the characteristics of the energy distribution remain the same as in the DSA. 

\begin{figure} 
 \includegraphics[width=\columnwidth]{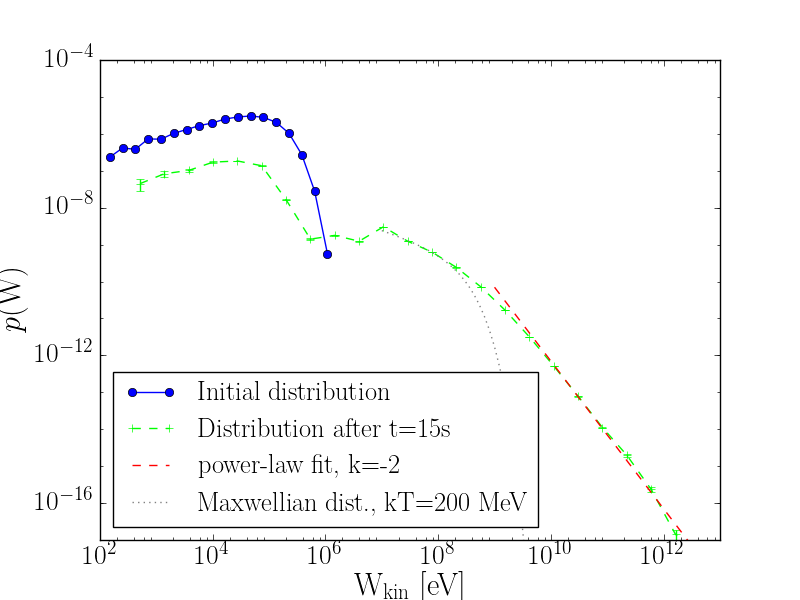}
 \caption{Ions in an environment of DSA combined with stochastic acceleration: Initial distribution (blue) and distribution of the ions escaping from the box (green) after $15$ seconds, with a power-law fit $\propto W^k$ with $k=-2$ (red), and a Maxwelian distribution with temperature $T= 200$ $GeV$ (gray).}
 \label{fig:dist_mix_ions}
\end{figure}

Ions in this mixed environment of stochastic acceleration and DSA  in an open simulation box  have approximately  the same $t_{esc}$ and $t_{acc}$ as the electrons. 
The kinetic energy distribution in the case of ions is shown in Figure~\ref{fig:dist_mix_ions}, it exhibits the same hybrid form of a Maxwellian and a power law  tail with index $k\approx-2$. The difference with the electrons is that the temperature of the ions is higher,  $T\approx 200\,$MeV, and the power-law tail reaches a higher maximum energy in the asymptotic state. 
$W_{shock}$ and $W_{scat}$ have the same relative contribution as previously reported for the electrons, so again the scattering process dominates. In the case of ions, the ratio $\left\langle W_{scat} \right\rangle / \left\langle W_{shock} \right\rangle$ is $\approx 5$, so it is slightly lower than in the case of electrons. The distribution of $W_{scat}$ is still a synthesis of hot plasma, but now with temperature $T\approx 200\,$MeV, which again coincides with the low energy part of the final total distribution, and a power law tail with index $k\approx -2$, while $W_{shock}$ has the same general form but with a temperature $T\approx 80\,$MeV in the Maxwellian low energy part.

\subsection{Diffusive Shock Acceleration  in the Presence of  Turbulent Reconnection}


The presence of large amplitude MHD disturbances, which is necessary to enhance the trapping of the charged particles in the vicinity of the shock, will spontaneously drive the TR state, as it is well documented in the literature reported in the introduction. 

\cite{Pisokas18} analyzed the synergy of stochastic and systematic acceleration that takes place when an unstable plasma reaches the state of turbulent reconnection. In this article, we expand many of the concepts developed by \cite{Pisokas18} in order to couple the Diffusive Shock Accelerator with Turbulent Acceleration in the vicinity of a shock.

\begin{figure} 
 \includegraphics[width=\columnwidth]{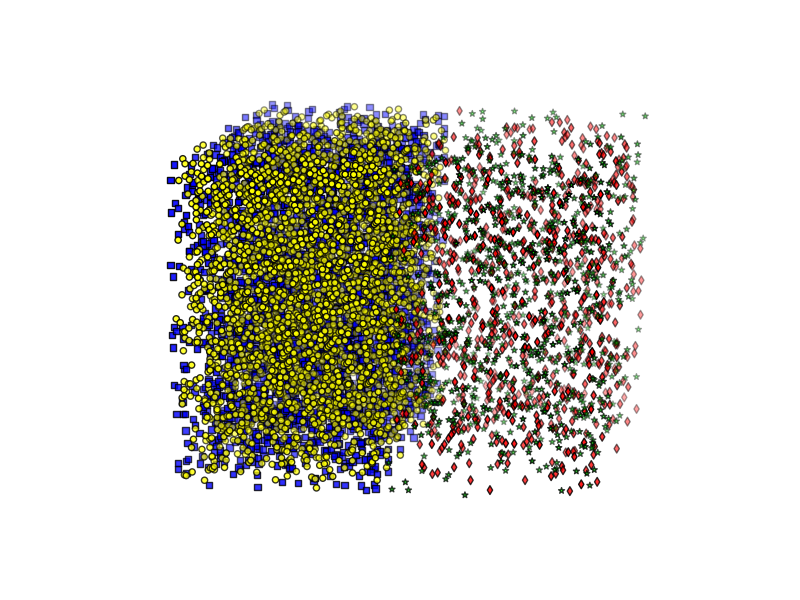}
 \caption{Small version of the box, with the stochastic scatterers upstream marked in red and downstream marked in blue. The randomly placed current sheets are marked in green upstream and in yellow downstream. 10\% of the grid points are active downstream (of which 50\% are stochastic scatterers and 50\% current sheets) and 2.5\% upstream (of which 50\% are stochastic scatterers and 50\% current sheets).}
 \label{fig:box3}
\end{figure}

We keep the same basic configuration for the simulation box, changing though the properties of the active grid points. Now the scatterers are divided into two classes, a fraction $P$ are stochastic scatterers, 
exactly as the ones considered in the previous subsection, and the rest ($1-P$) are unstable current sheets (UCSs), see \cite{Pisokas18}. For the rest of this section we choose $P=0.5$, the environment we consider is illustrated in Fig.\ \ref{fig:box3}.
The injected particles have the same initial distribution 
as before. When a particle meets a large amplitude magnetic disturbance, its energy changes according to Eq.\ (\ref{eq:energy}), as before.
For UCSs as scatterers, the energy gain is caused  by electric fields and it follows the relation
\begin{equation}
\Delta W = \mid e \mid E_{eff} \ell_{eff} ,
\label{eq:current }
\end{equation}
with $e$ the charge of a particle, and $E_{eff} \approx (U_{1/2}/c)\ \delta B$ is a measure of the effective electric field of the UCS \citep{Kowal11}. $U_{1}$ and $U_{2}$ are the velocity of the flow upstream and downstream, respectively, and $\delta B$ is the fluctuating magnetic field. Reconnection at an UCS induces stochastic fluctuations, so $\delta B$ is of stochastic nature, and we assume that it obeys a power law distribution with index $5/3$ in the range [$10^{-5} G$; $100 G$], i.e.\ it follows a Kolmogorov spectrum. Finally, the effective length $\ell_{eff}$ is assumed as an increasing linear function of $E_{eff}$, restricted to values between $10\,$m and $1\,$km. We note that the interaction between particles and UCSs always leads to a positive energy increment (systematic energy gain), which also is not dependent on the instantaneous energy of the particles.
The synergy of large scale magnetic disturbances with randomly distributed UCS, studied in \cite{Pisokas18}, is now investigated in the presence of an additional systematic accelerator, the DSA.

\begin{figure*} 
(a)
 \includegraphics[width=\columnwidth]{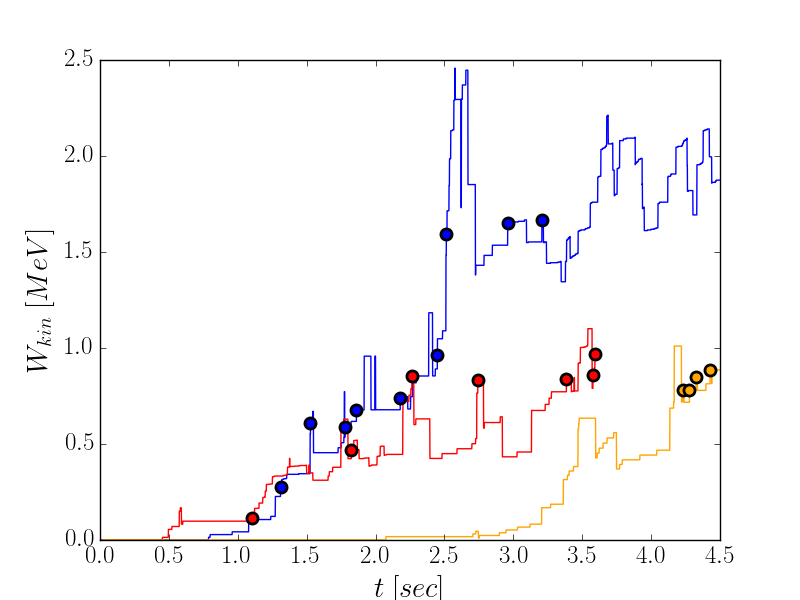}
 (b)
 \includegraphics[width=\columnwidth]{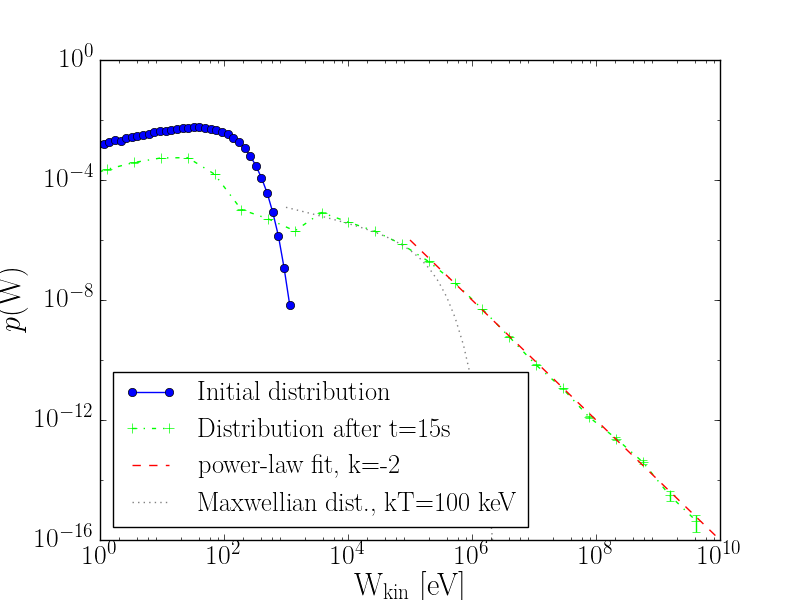}
 (c)
 \includegraphics[width=\columnwidth]{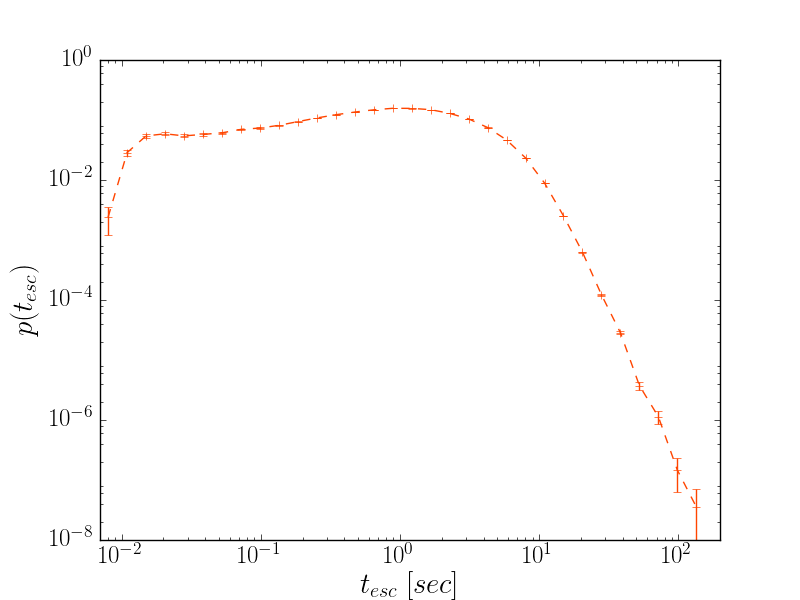}
 (d)
 \includegraphics[width=\columnwidth]
 {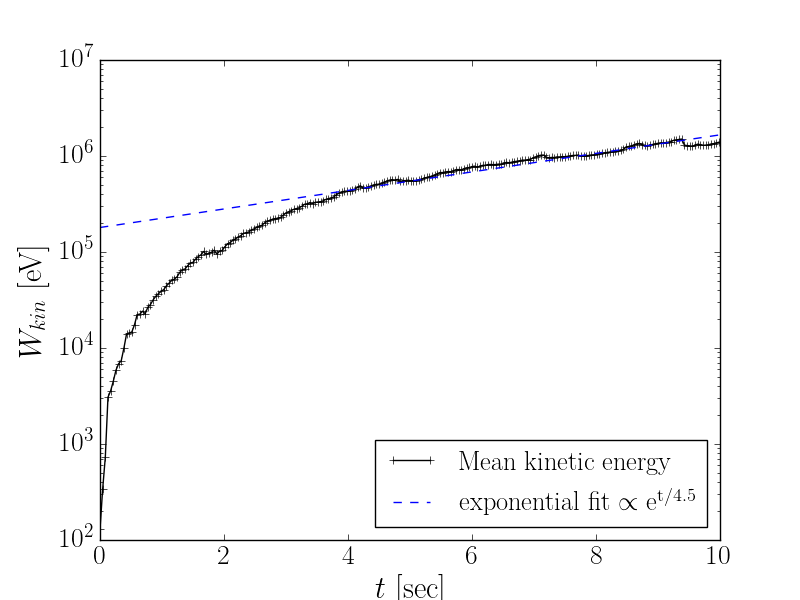}
(e)
 \includegraphics[width=\columnwidth]{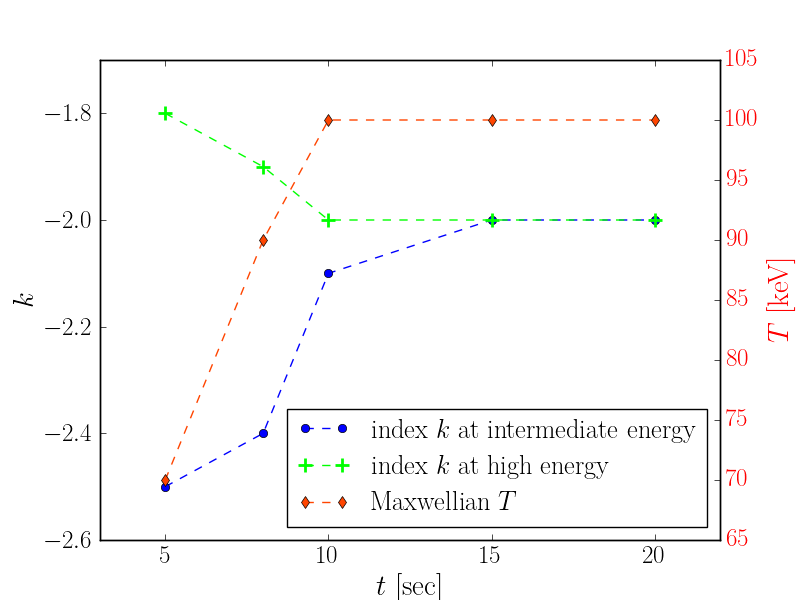}
(f)
 \includegraphics[width=\columnwidth]{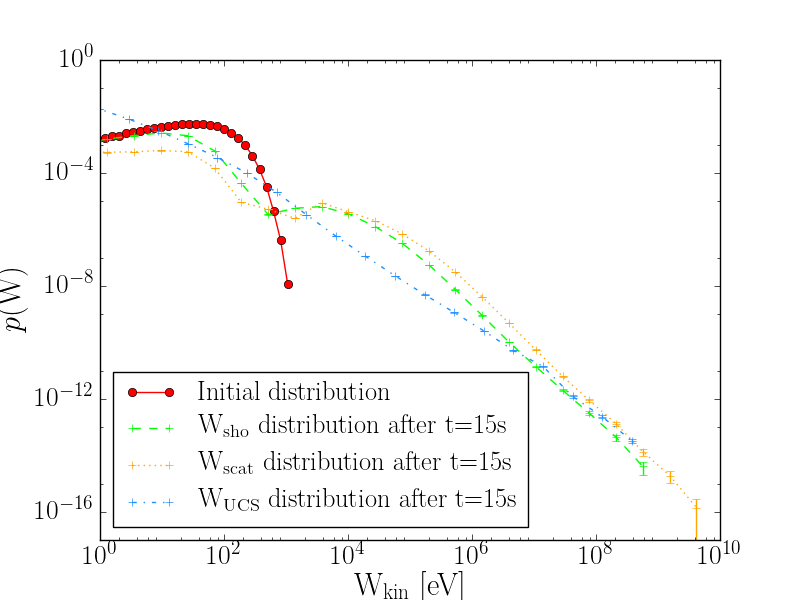}
 \caption{Acceleration of electrons by the synergy of turbulent reconnection and DSA: 
(a) The kinetic energy evolution of typical electrons interacting with the three accelerators. The circles mark the crossings of the shock surface.
(b)  Initial kinetic energy distribution (blue) and distribution after 15 seconds for electrons escaping from the box (green), with a power-law fit ($= W^{-2}$ (red)), and a Maxwellian distribution with temperature $T= 100$ $keV$ (gray). 
 (c) Distribution of the escape times $t_{esc}$. 
 (d) Mean kinetic energy as a function of time (black), with an exponential fit (blue). 
 (e) Evolution of the index $k$ of the power-law tail at high energies (green), and of the power-law index at intermediate energies (blue),  together with the evolution of the temperature $T$ of the Maxwellian fit (red). 
 (f) Initial kinetic energy distribution (red), the distribution from the shock contribution only (green), from the stochastic scatterers contribution only (orange), and from the UCSs contribution only (blue).}
 \label{fig:3part}
\end{figure*}

Figure~\ref{fig:3part}(a) shows the time evolution of the kinetic energy of a few typical particles. The picture here is similar to the case of DSA together with stochastic scatterers, 
Fig.\ \ref{fig:mix}(a), there is a random walk like behavior with an increasing trend, the latter being though more pronounced in the presence of UCSs that are also systematic accelerators.

The energy distribution of the electrons escaping from the simulation box after reaching the asymptotic state  is again a synthesis of a hot plasma with temperature $\approx 100$ $keV$ and a high energy power-law tail with index $k\approx-2$ (see Figure~\ref{fig:3part}(b)). The maximum energy particles reach is $\approx 10$ $GeV$, so it is comparable with the case we discussed before where UCS were absent ($P=1$). 	We note that at times before $t=15\,$sec, there actually is a double power-law, one at intermediate energies and one at the high energy tail, which at about $t=15\,$sec merge and attain to the same index -2, see Figure~\ref{fig:3part}(e). The appearance of this double power law must be attributed to the presence of the UCSs. The UCSs' effect is thus of equal importance than the one of the DSA and the stochastic scatterers, mainly though at early times, before the asymptotic state is being reached.
	That the action of the UCSs is limited to early times can be explained by the fact that in their case the energy increments do not depend on the energy of the particles, so they are most efficient in the phase where particles still have relatively low energies, and at later times the more energetic particles are accelerated mostly by the systematic acceleration of the DSA and by the stochastic acceleration through the large amplitude magnetic disturbances.

The median value of the escape times is $t_{esc}\approx 3.7\,$sec (see Fig. (\ref{fig:3part}(c)), so it is about one second larger than in the case of $P=1$. This result is in agreement with the result reported by \cite{Pisokas18} on the role played by the UCS in the trapping of  particles inside a Turbulently Reconnecting volume. The acceleration time is increased compared to the $P=1$ case, we find $t_{acc}\approx 4.5$ seconds (see Figure~\ref{fig:3part}(d)).

Keeping track of the source of each energy gain from one of the three acceleration processes (stochastic acceleration by large scale magnetic disturbances, acceleration at UCS, and DSA) along the trajectories of the particles, we can analyze the contribution of each accelerator separately. 
Figure~\ref{fig:3part}(f) shows the kinetic energy distribution due to each acceleration process separately. In all cases there are power-law tails, with the DSA and the stochastic scatterers exhibiting the same index -2, and the power-law due to the UCSs being flatter with index -1.4 at the intermediate energies and index -1.8 at the high energies. This again confirms that at large times the role of the UCSs is limited.


It is important to note that the synergy of the UCS and the DSA (the case $P=0$) has been studied by several authors (see \cite{Zank15, leroux16}), who pointed out that the energy distribution of the accelerated particles is harder ($\sim 1.8$) in this case. As we already mentioned, this result also appears in our simulation at times before the energy distribution reaches an asymptotic state (see Fig. ~\ref{fig:3part}(e) at about $t=5$ sec). Our study thus lets us conclude that the index 1.8 found by \cite{Zank15, leroux16} is the result of the action of UCSs, and not of DSA.

The asymptotic energy distribution of the ions is slightly different from the one of the electrons, the power-law tail index is now $k\approx -2.2$, coinciding with the one following from the relation $k= -(1+ t_{acc}/t_{esc})$, with the same $t_{acc}$ ($\approx 4.5\,$sec) as for $P=1$ (DSA together with stochastic scatterers) and also almost the same $t_{esc}$ ($\approx 3.8$ sec), and thus, with the prediction of Eq.\ (\ref{eq:imp}) being fulfilled, we can say that stochastic acceleration together with DSA dominates the energization of the ions. Also, the ions reach a higher maximal energy ($\approx 1000$ $GeV$), and from a Maxwellian fit to the low energy particles' distribution we recover the fact that ions are hotter than electrons ($T\approx 200$ $MeV$). 

\section{Discussion and summary}

In this article we simulate a large scale shock in the upper solar corona, formed e.g.\ by a Coronal Mass Ejection, till the energy distributions of electrons and ions reach their asymptotic state. We use as many pieces of information as possible from the microscopic analysis of current numerical simulations, but our emphasis here is on realistic spatial and temporal scales. We move progressively from elastic scatterers to active scatterers (stochastic interaction with large amplitude magnetic fluctuations and systematic scattering at UCSs) in the vicinity of the shock, and 
our main results are:

\begin{enumerate}
	\item We have analyzed three distinct accelerators (1) DSA with passive scatterers, (2) DSA with stochastic scatterers, (3) DSA with turbulent reconnection (a combination of stochastic scatterers, as before, and UCSs). The asymptotic energy distribution of the particles escaping from the simulation box  remains the same in all numerical experiments, a mixture of a hot plasma with a power law tail with index -2 for the high energy particles. 
	\item The presence of active scatterers in the vicinity of the shock does not affect the universal form of the energy distribution, only the acceleration time becomes shorter  and the maximum energy reached by the high energy particles increases by one or two orders of magnitude. 
	\item Keeping track of the source of the energy gain along the particles' trajectories, we have shown that the active scatterers contribute 5 to 6 times more energy than the DSA, and the energy distribution of the particles retains the main characteristics we have found in the case of DSA with passive scatterers. 
	\item The synergy of DSA and UCSs gives the same mixture of a hot plasma and a power law tail, yet with harder  power law index ($ k \sim 1.8)$ at early times, as it had been pointed out already in the literature (see \cite{Zank15, leroux16}).
	\item The appearance of a power law tail with index -2 often is considered to be characteristic for DSA, we though stress here that this result has been found for TR alone, in the absence of a shock, see \cite{Pisokas18}. TR thus gives results very similar with the ones of the synergy of DSA and TR.
	\item The ions acquire the same shape of the energy distribution as the electrons, yet their energy gain is on slower scales, the maximum energy reached is higher than for the electrons, and the temperature of the low energy particles is higher than the one of the electrons.
\end{enumerate}

In our opinion, at a traveling shock (e.g.\ during a Coronal Mass Ejection) initially DSA may dominate the heating and acceleration of particles, soon though the synergy of DSA and TR will become the main acceleration mechanism. In the final stages, when the nonlinear phenomena have become very strong and the shock has been fragmented, TR continues as a moving sheath around the shock surface. The remarkable result from our study is that in all these stages of the shock evolution the accelerated particles have  a universal energy distribution in the asymptotic state, namely a synthesis of a Maxwellian for the low energy particles with a power law tail with constant index for the high energy particles.

\section*{Acknowledgements}
C. Garrel thanks the Observatoire de la $C\hat{o}te \; d'  Azur$, the University of Thessaloniki and the Erasmus$^+$ program for making this joint research work possible.
L. Vlahos and H. Isliker thank  the national Programme for the Controlled Thermonuclear Fusion, Hellenic Republic for their financial support. The sponsors do not bear any responsibility for the content of this work.


\begin{thebibliography}{}
\makeatletter
\relax
\def\mn@urlcharsother{\let\do\@makeother \do\$\do\&\do\#\do\^\do\_\do\%\do\~}
\def\mn@doi{\begingroup\mn@urlcharsother \@ifnextchar [ {\mn@doi@}
  {\mn@doi@[]}}
\def\mn@doi@[#1]#2{\def\@tempa{#1}\ifx\@tempa\@empty \href
  {http://dx.doi.org/#2} {doi:#2}\else \href {http://dx.doi.org/#2} {#1}\fi
  \endgroup}
\def\mn@eprint#1#2{\mn@eprint@#1:#2::\@nil}
\def\mn@eprint@arXiv#1{\href {http://arxiv.org/abs/#1} {{\tt arXiv:#1}}}
\def\mn@eprint@dblp#1{\href {http://dblp.uni-trier.de/rec/bibtex/#1.xml}
  {dblp:#1}}
\def\mn@eprint@#1:#2:#3:#4\@nil{\def\@tempa {#1}\def\@tempb {#2}\def\@tempc
  {#3}\ifx \@tempc \@empty \let \@tempc \@tempb \let \@tempb \@tempa \fi \ifx
  \@tempb \@empty \def\@tempb {arXiv}\fi \@ifundefined
  {mn@eprint@\@tempb}{\@tempb:\@tempc}{\expandafter \expandafter \csname
  mn@eprint@\@tempb\endcsname \expandafter{\@tempc}}}

\bibitem[\protect\citeauthoryear{{Achterberg}}{{Achterberg}}{1981}]{Achterberg81}
{Achterberg} A.,  1981, \aap, \href
  {http://adsabs.harvard.edu/abs/1981A%26A....97..259A} {97, 259}

\bibitem[\protect\citeauthoryear{{Achterberg}}{{Achterberg}}{2008}]{Achterberg08}
{Achterberg} A.,  2008, Cosmic Accelerators, Lecture Notes for the Inter-University Lectures Series 
\bibitem[\protect\citeauthoryear{Ambrosiano, Matthaeus, Goldstein  \&
  Plante}{Ambrosiano et~al.}{1988}]{Ambrosiano88}
Ambrosiano J.,  Matthaeus W.~H.,  Goldstein M.~L.,   Plante D.,  1988, Journal
  of Geophysical Research: Space Physics, 93, 14383

\bibitem[\protect\citeauthoryear{Arzner \& Vlahos}{Arzner \&
  Vlahos}{2004}]{Arzner04}
Arzner K.,  Vlahos L.,  2004, Astrophys. J. Lett., 605, L69

\bibitem[\protect\citeauthoryear{{Arzner}, {Knaepen}, {Carati}, {Denewet}  \&
  {Vlahos}}{{Arzner} et~al.}{2006}]{Arzner06}
{Arzner} K.,  {Knaepen} B.,  {Carati} D.,  {Denewet} N.,   {Vlahos} L.,  2006,
  \mn@doi [Astrophys. J.] {10.1086/498341}, \href
  {http://adsabs.harvard.edu/abs/2006ApJ...637..322A} {637, 322}

\bibitem[\protect\citeauthoryear{{Axford}, {Leer}  \& {Skadron}}{{Axford}
  et~al.}{1978}]{Axford78}
{Axford} W.~I.,  {Leer} E.,   {Skadron} G.,  1978, in {Shaviv} G.,  ed.,
  Cosmophysics. pp 125--134

\bibitem[\protect\citeauthoryear{{Bai}, {Caprioli}, {Sironi}  \&
  {Spitkovsky}}{{Bai} et~al.}{2015}]{Bai15}
{Bai} X.-N.,  {Caprioli} D.,  {Sironi} L.,   {Spitkovsky} A.,  2015, \mn@doi
  [\apj] {10.1088/0004-637X/809/1/55}, \href
  {http://adsabs.harvard.edu/abs/2015ApJ...809...55B} {809, 55}

\bibitem[\protect\citeauthoryear{Balsara,   Benjamin \& Cox}{Balsara
  et~al.}{2001}]{Balsara01}
Balsara D.,  Benjamin R.,   Cox D.~P.,  2001, \apj, 563, 800

\bibitem[\protect\citeauthoryear{Balsara,   \& Kim}{Balsara
  \& Kim}{2005}]{Balsara05}
Balsara D.,  Kim J.,  2005, \apj, 634, 390


\bibitem[\protect\citeauthoryear{{Bell}}{{Bell}}{1978}]{Bell78a}
{Bell} A.~R.,  1978, \mnras, \href
  {http://adsabs.harvard.edu/abs/1978MNRAS.182..147B} {182, 147}
  
\bibitem[\protect\citeauthoryear{{Bell}}{{Bell}}{2004}]{Bell04}
{Bell} A.~R.,  2004,  \mnras, 353, 550

\bibitem[\protect\citeauthoryear{{Biskamp} \& {Welter}}{{Biskamp} \&
  {Welter}}{1989}]{Biskamp89}
{Biskamp} D.,  {Welter} H.,  1989, \mn@doi [Physics of Fluids B]
  {10.1063/1.859060}, \href {http://adsabs.harvard.edu/abs/1989PhFlB...1.1964B}
  {1, 1964}

\bibitem[\protect\citeauthoryear{{Blandford} \& {Ostriker}}{{Blandford} \&
  {Ostriker}}{1978}]{Blandford78}
{Blandford} R.~D.,  {Ostriker} J.~P.,  1978, \mn@doi [\apjl] {10.1086/182658},
  \href {http://adsabs.harvard.edu/abs/1978ApJ...221L..29B} {221, L29}

\bibitem[\protect\citeauthoryear{{Burgess}, {M{\"o}bius}  \&
  {Scholer}}{{Burgess} et~al.}{2012}]{Burgess12}
{Burgess} D.,  {M{\"o}bius} E.,   {Scholer} M.,  2012, \mn@doi [\ssr]
  {10.1007/s11214-012-9901-5}, \href
  {http://adsabs.harvard.edu/abs/2012SSRv..173....5B} {173, 5}

\bibitem[\protect\citeauthoryear{{Burgess}, {Gingell}  \& {Matteini}}{{Burgess}
  et~al.}{2016}]{Burgess16}
{Burgess} D.,  {Gingell} P.~W.,   {Matteini} L.,  2016, \mn@doi [Astrophys. J.]
  {10.3847/0004-637X/822/1/38}, \href
  {http://adsabs.harvard.edu/abs/2016ApJ...822...38B} {822, 38}

\bibitem[\protect\citeauthoryear{{Bykov}, {Brandenburg}, {Malkov}  \&
  {Osipov}}{{Bykov} et~al.}{2013}]{Bykov13}
{Bykov} A.~M.,  {Brandenburg} A.,  {Malkov} M.~A.,   {Osipov} S.~M.,  2013,
  \mn@doi [\ssr] {10.1007/s11214-013-9988-3}, \href
  {http://adsabs.harvard.edu/abs/2013SSRv..178..201B} {178, 201}

\bibitem[\protect\citeauthoryear{{Caprioli} \& {Spitkovsky}}{{Caprioli} \&
  {Spitkovsky}}{2014a}]{Caprioli14a}
{Caprioli} D.,  {Spitkovsky} A.,  2014a, \mn@doi [\apj]
  {10.1088/0004-637X/783/2/91}, \href
  {http://adsabs.harvard.edu/abs/2014ApJ...783...91C} {783, 91}

\bibitem[\protect\citeauthoryear{{Caprioli} \& {Spitkovsky}}{{Caprioli} \&
  {Spitkovsky}}{2014b}]{Caprioli14b}
{Caprioli} D.,  {Spitkovsky} A.,  2014b, \mn@doi [\apj]
  {10.1088/0004-637X/794/1/46}, \href
  {http://adsabs.harvard.edu/abs/2014ApJ...794...46C} {794, 46}

\bibitem[\protect\citeauthoryear{{Caprioli} \& {Spitkovsky}}{{Caprioli} \&
  {Spitkovsky}}{2014c}]{Caprioli14c}
{Caprioli} D.,  {Spitkovsky} A.,  2014c, \mn@doi [\apj]
  {10.1088/0004-637X/794/1/47}, \href
  {http://adsabs.harvard.edu/abs/2014ApJ...794...47C} {794, 47}

\bibitem[\protect\citeauthoryear{{Chasapis} et~al.,}{{Chasapis}
  et~al.}{2015}]{Chasapis15}
{Chasapis} A.,  et~al., 2015, \mn@doi [\apjl] {10.1088/2041-8205/804/1/L1},
  \href {http://adsabs.harvard.edu/abs/2015ApJ...804L...1C} {804, L1}
  
\bibitem[\protect\citeauthoryear{del Valle, Lazarian  \& Santos-Lima}{del Valle
  et~al.}{2018}]{delVale16}
del Valle, M.V.,  Lazarian A.,   Santos-Lima, R,,  2016, \mnras, 458, 1645


\bibitem[\protect\citeauthoryear{Dmitruk, Matthaeus, Seenu  \& Brown}{Dmitruk
  et~al.}{2003}]{Dmitruk03}
Dmitruk P.,  Matthaeus W.,  Seenu N.,   Brown M.~R.,  2003, The Astrophysical
  Journal Letters, 597, L81

\bibitem[\protect\citeauthoryear{{Dmitruk}, {Matthaeus}  \& {Seenu}}{{Dmitruk}
  et~al.}{2004}]{Dmitruk04}
{Dmitruk} P.,  {Matthaeus} W.~H.,   {Seenu} N.,  2004, \mn@doi [Astrophys. J.]
  {10.1086/425301}, \href {http://adsabs.harvard.edu/abs/2004ApJ...617..667D}
  {617, 667}

\bibitem[\protect\citeauthoryear{{Drake}, {Opher}, {Swisdak}  \&
  {Chamoun}}{{Drake} et~al.}{2010}]{Drake10}
{Drake} J.~F.,  {Opher} M.,  {Swisdak} M.,   {Chamoun} J.~N.,  2010, \mn@doi
  [Astrophys. J.] {10.1088/0004-637X/709/2/963}, \href
  {http://adsabs.harvard.edu/abs/2010ApJ...709..963D} {709, 963}

\bibitem[\protect\citeauthoryear{{Drury}}{{Drury}}{1983}]{Drury83}
{Drury} L.~O.,  1983, \mn@doi [Reports on Progress in Physics]
  {10.1088/0034-4885/46/8/002}, \href
  {http://adsabs.harvard.edu/abs/1983RPPh...46..973D} {46, 973}

\bibitem[\protect\citeauthoryear{{Fermi}}{{Fermi}}{1949}]{Fermi49}
{Fermi} E.,  1949, \mn@doi [Physical Review] {10.1103/PhysRev.75.1169}, \href
  {http://adsabs.harvard.edu/abs/1949PhRv...75.1169F} {75, 1169}

\bibitem[\protect\citeauthoryear{{Fermi}}{{Fermi}}{1954}]{Fermi54}
{Fermi} E.,  1954, \mn@doi [The Astrophysical Journal] {10.1086/145789}, \href
  {http://adsabs.harvard.edu/abs/1954ApJ...119....1F} {119, 1}

\bibitem[\protect\citeauthoryear{Giacalone  \& Jokipii}{Giacalone \& Jokipii}{2007}]{Giacalone07}
Giacalone J.,  Jokipii J.R.,  2018, \apj, 663, L41


\bibitem[\protect\citeauthoryear{{Isliker}, {Vlahos}  \&
  {Constantinescu}}{{Isliker} et~al.}{2017a}]{Isliker17a}
{Isliker} H.,  {Vlahos} L.,   {Constantinescu} D.,  2017a, \mn@doi [Physical
  Review Letters] {10.1103/PhysRevLett.119.045101}, \href
  {http://adsabs.harvard.edu/abs/2017PhRvL.119d5101I} {119, 045101}

\bibitem[\protect\citeauthoryear{{Isliker}, {Pisokas}, {Vlahos}  \&
  {Anastasiadis}}{{Isliker} et~al.}{2017b}]{Isliker17}
{Isliker} H.,  {Pisokas} T.,  {Vlahos} L.,   {Anastasiadis} A.,  2017b, \mn@doi
  [\apj] {10.3847/1538-4357/aa8ee8}, \href
  {http://adsabs.harvard.edu/abs/2017ApJ...849...35I} {849, 35}

\bibitem[\protect\citeauthoryear{{Jones}}{{Jones}}{1994}]{Jones94}
{Jones} F.~C.,  1994, \mn@doi [\apjs] {10.1086/191875}, \href
  {http://adsabs.harvard.edu/abs/1994ApJS...90..561J} {90, 561}

\bibitem[\protect\citeauthoryear{{Karimabadi} \& {Lazarian}}{{Karimabadi} \&
  {Lazarian}}{2013}]{Karibabadi2013c}
{Karimabadi} H.,  {Lazarian} A.,  2013, \mn@doi [Physics of Plasmas]
  {10.1063/1.4828395}, \href
  {http://adsabs.harvard.edu/abs/2013PhPl...20k2102K} {20, 112102}

\bibitem[\protect\citeauthoryear{{Karimabadi} et~al.,}{{Karimabadi}
  et~al.}{2014}]{Karimabadi2014}
{Karimabadi} H.,  et~al., 2014, \mn@doi [Physics of Plasmas]
  {10.1063/1.4882875}, \href
  {http://adsabs.harvard.edu/abs/2014PhPl...21f2308K} {21, 062308}

\bibitem[\protect\citeauthoryear{Karimabadi, Roytershteyn, Daughton  \&
  Liu}{Karimabadi et~al.}{013a}]{Karimabadi13a}
Karimabadi H.,  Roytershteyn V.,  Daughton W.,   Liu Y.-H.,  2013a, Space
  Science Reviews, 178, 307

\bibitem[\protect\citeauthoryear{{Khabarova} \& {Zank}}{{Khabarova} \&
  {Zank}}{2017}]{Khabarova17}
{Khabarova} O.~V.,  {Zank} G.~P.,  2017, \mn@doi [\apj]
  {10.3847/1538-4357/aa7686}, \href
  {http://adsabs.harvard.edu/abs/2017ApJ...843....4K} {843, 4}

\bibitem[\protect\citeauthoryear{{Khabarova}, {Zank}, {Li}, {Malandraki}, {le
  Roux}  \& {Webb}}{{Khabarova} et~al.}{2016}]{Khabarova16}
{Khabarova} O.~V.,  {Zank} G.~P.,  {Li} G.,  {Malandraki} O.~E.,  {le Roux}
  J.~A.,   {Webb} G.~M.,  2016, \mn@doi [\apj] {10.3847/0004-637X/827/2/122},
  \href {http://adsabs.harvard.edu/abs/2016ApJ...827..122K} {827, 122}

\bibitem[\protect\citeauthoryear{{Kowal}, {de Gouveia Dal Pino}  \&
  {Lazarian}}{{Kowal} et~al.}{2011}]{Kowal11}
{Kowal} G.,  {de Gouveia Dal Pino} E.~M.,   {Lazarian} A.,  2011, \mn@doi
  [\apj] {10.1088/0004-637X/735/2/102}, \href
  {http://adsabs.harvard.edu/abs/2011ApJ...735..102K} {735, 102}

\bibitem[\protect\citeauthoryear{{Krymskii}}{{Krymskii}}{1977}]{Krymskii77}
{Krymskii} G.~F.,  1977, Akademiia Nauk SSSR Doklady, \href
  {http://adsabs.harvard.edu/abs/1977DoSSR.234.1306K} {234, 1306}

\bibitem[\protect\citeauthoryear{Kulsrud \& Ferrari}{Kulsrud \&
  Ferrari}{1971}]{Kulsrud71}
Kulsrud R.~M.,  Ferrari A.,  1971, Astrophysics and Space Science, 12, 302

\bibitem[\protect\citeauthoryear{{Lazarian} \& {Opher}}{{Lazarian} \&
  {Opher}}{2009}]{Lazarian09}
{Lazarian} A.,  {Opher} M.,  2009, \mn@doi [\apj] {10.1088/0004-637X/703/1/8},
  \href {http://adsabs.harvard.edu/abs/2009ApJ...703....8L} {703, 8}

\bibitem[\protect\citeauthoryear{{Lazarian} \& {Vishniac}}{{Lazarian} \&
  {Vishniac}}{1999}]{Lazarian99}
{Lazarian} A.,  {Vishniac} E.~T.,  1999, \mn@doi [Astrophys. J.]
  {10.1086/307233}, \href {http://adsabs.harvard.edu/abs/1999ApJ...517..700L}
  {517, 700}

\bibitem[\protect\citeauthoryear{{Lazarian}, {Eyink}, {Vishniac}  \&
  {Kowal}}{{Lazarian} et~al.}{2015}]{Lazarian15}
{Lazarian} A.,  {Eyink} G.,  {Vishniac} E.,   {Kowal} G.,  2015, \mn@doi
  [Philosophical Transactions of the Royal Society of London Series A]
  {10.1098/rsta.2014.0144}, \href
  {http://adsabs.harvard.edu/abs/2015RSPTA.37340144L} {373, 20140144}

\bibitem[\protect\citeauthoryear{{Longair}}{{Longair}}{2011}]{Longair11}
{Longair} M.~S.,  2011, {High Energy Astrophysics}.
Cambridge University Press

\bibitem[\protect\citeauthoryear{{Matsumoto}, {Amano}, {Kato}  \&
  {Hoshino}}{{Matsumoto} et~al.}{2015}]{Matsumoto15}
{Matsumoto} Y.,  {Amano} T.,  {Kato} T.~N.,   {Hoshino} M.,  2015, \mn@doi
  [Science] {10.1126/science.1260168}, \href
  {http://adsabs.harvard.edu/abs/2015Sci...347..974M} {347, 974}

\bibitem[\protect\citeauthoryear{{Matthaeus} \& {Lamkin}}{{Matthaeus} \&
  {Lamkin}}{1986}]{Matthaeus86}
{Matthaeus} W.~H.,  {Lamkin} S.~L.,  1986, \mn@doi [Physics of Fluids]
  {10.1063/1.866004}, \href {http://adsabs.harvard.edu/abs/1986PhFl...29.2513M}
  {29, 2513}

\bibitem[\protect\citeauthoryear{{Matthaeus} \& {Velli}}{{Matthaeus} \&
  {Velli}}{2011}]{Matthaeus11}
{Matthaeus} W.~H.,  {Velli} M.,  2011, \mn@doi [Space Science Reviews]
  {10.1007/s11214-011-9793-9}, \href
  {http://adsabs.harvard.edu/abs/2011SSRv..160..145M} {160, 145}

\bibitem[\protect\citeauthoryear{Matthaeus, Wan, Servidio, Greco, Osman,
  Oughton  \& Dmitruk}{Matthaeus et~al.}{2015}]{Matthaeus15}
Matthaeus W.~H.,  Wan M.,  Servidio S.,  Greco A.,  Osman K.~T.,  Oughton S.,
  Dmitruk P.,  2015, \mn@doi [Philosophical Transactions of the Royal Society
  of London A: Mathematical, Physical and Engineering Sciences]
  {10.1098/rsta.2014.0154}, 373

\bibitem[\protect\citeauthoryear{{Melrose}}{{Melrose}}{1994}]{Melrose94}
{Melrose} D.~B.,  1994, \mn@doi [\apjs] {10.1086/191883}, \href
  {http://adsabs.harvard.edu/abs/1994ApJS...90..623M} {90, 623}

\bibitem[\protect\citeauthoryear{{Melrose}}{{Melrose}}{2009}]{Melrose2009}
{Melrose} D.~B.,  2009, preprint, \href
  {http://adsabs.harvard.edu/abs/2009arXiv0902.1803M} {} (\mn@eprint {arXiv}
  {0902.1803})

\bibitem[\protect\citeauthoryear{{Miller} \& {Ramaty}}{{Miller} \&
  {Ramaty}}{1992}]{Miller92}
{Miller} J.~A.,  {Ramaty} R.,  1992, in {Zank} G.~P.,  {Gaisser} T.~K.,  eds,
  American Institute of Physics Conference Series Vol. 264, Particle
  Acceleration in Cosmic Plasmas. pp 223--228, \mn@doi{10.1063/1.42732}

\bibitem[\protect\citeauthoryear{{Miller}, {Guessoum}  \& {Ramaty}}{{Miller}
  et~al.}{1990}]{Miller90}
{Miller} J.~A.,  {Guessoum} N.,   {Ramaty} R.,  1990, \mn@doi [\apj]
  {10.1086/169233}, \href {http://adsabs.harvard.edu/abs/1990ApJ...361..701M}
  {361, 701}

\bibitem[\protect\citeauthoryear{{Miller} et~al.,}{{Miller}
  et~al.}{1997}]{Miller97}
{Miller} J.~A.,  et~al., 1997, \mn@doi [\jgr] {10.1029/97JA00976}, \href
  {http://adsabs.harvard.edu/abs/1997JGR...10214631M} {102, 14631}

\bibitem[\protect\citeauthoryear{{Osman}, {Matthaeus}, {Gosling}, {Greco},
  {Servidio}, {Hnat}, {Chapman}  \& {Phan}}{{Osman} et~al.}{2014}]{Osman14}
{Osman} K.~T.,  {Matthaeus} W.~H.,  {Gosling} J.~T.,  {Greco} A.,  {Servidio}
  S.,  {Hnat} B.,  {Chapman} S.~C.,   {Phan} T.~D.,  2014, \mn@doi [Physical
  Review Letters] {10.1103/PhysRevLett.112.215002}, \href
  {http://adsabs.harvard.edu/abs/2014PhRvL.112u5002O} {112, 215002}

\bibitem[\protect\citeauthoryear{{Parker} \& {Tidman}}{{Parker} \&
  {Tidman}}{1958}]{Parker58}
{Parker} E.~N.,  {Tidman} D.~A.,  1958, \mn@doi [Physical Review]
  {10.1103/PhysRev.111.1206}, \href
  {http://adsabs.harvard.edu/abs/1958PhRv..111.1206P} {111, 1206}

\bibitem[\protect\citeauthoryear{Petrosian}{Petrosian}{2012}]{Petrosian12}
Petrosian V.,  2012, Space science reviews, 173, 535

\bibitem[\protect\citeauthoryear{{Pisokas}, {Vlahos}, {Isliker}, {Tsiolis}  \&
  {Anastasiadis}}{{Pisokas} et~al.}{2017}]{Pisokas17}
{Pisokas} T.,  {Vlahos} L.,  {Isliker} H.,  {Tsiolis} V.,   {Anastasiadis} A.,
  2017, \mn@doi [\apj] {10.3847/1538-4357/835/2/214}, \href
  {http://adsabs.harvard.edu/abs/2017ApJ...835..214P} {835, 214}

\bibitem[\protect\citeauthoryear{Pisokas, Vlahos  \& Isliker}{Pisokas
  et~al.}{2018}]{Pisokas18}
Pisokas T.,  Vlahos L.,   Isliker H.,  2018, \apj, 852, 64

\bibitem[\protect\citeauthoryear{{Ramaty}}{{Ramaty}}{1979}]{Ramaty79}
{Ramaty} R.,  1979, in {Arons} J.,  {McKee} C.,   {Max} C.,  eds,  American
  Institute of Physics Conference Series Vol. 56, Particle Acceleration
  Mechanisms in Astrophysics. pp 135--154, \mn@doi{10.1063/1.32074}

\bibitem[\protect\citeauthoryear{{Schure}, {Bell}, {O'C Drury}  \&
  {Bykov}}{{Schure} et~al.}{2012}]{Schure12}
{Schure} K.~M.,  {Bell} A.~R.,  {O'C Drury} L.,   {Bykov} A.~M.,  2012, \mn@doi
  [\ssr] {10.1007/s11214-012-9871-7}, \href
  {http://adsabs.harvard.edu/abs/2012SSRv..173..491S} {173, 491}

\bibitem[\protect\citeauthoryear{{Servidio}, {Matthaeus}, {Shay}, {Dmitruk},
  {Cassak}  \& {Wan}}{{Servidio} et~al.}{2010}]{Servidio10}
{Servidio} S.,  {Matthaeus} W.~H.,  {Shay} M.~A.,  {Dmitruk} P.,  {Cassak}
  P.~A.,   {Wan} M.,  2010, \mn@doi [Physics of Plasmas] {10.1063/1.3368798},
  \href {http://adsabs.harvard.edu/abs/2010PhPl...17c2315S} {17, 032315}

\bibitem[\protect\citeauthoryear{{Servidio} et~al.,}{{Servidio}
  et~al.}{2011}]{Servidio11}
{Servidio} S.,  et~al., 2011, \mn@doi [Nonlinear Processes in Geophysics]
  {10.5194/npg-18-675-2011}, \href
  {http://adsabs.harvard.edu/abs/2011NPGeo..18..675S} {18, 675}

\bibitem[\protect\citeauthoryear{{Vlahos}, {Pisokas}, {Isliker}, {Tsiolis}  \&
  {Anastasiadis}}{{Vlahos} et~al.}{2016}]{Vlahos16}
{Vlahos} L.,  {Pisokas} T.,  {Isliker} H.,  {Tsiolis} V.,   {Anastasiadis} A.,
  2016, \mn@doi [Astrophys. J] {10.3847/2041-8205/827/1/L3}, \href
  {http://adsabs.harvard.edu/abs/2016ApJ...827L...3V} {827, L3}

\bibitem[\protect\citeauthoryear{{Zank}, {le Roux}, {Webb}, {Dosch}  \&
  {Khabarova}}{{Zank} et~al.}{2014}]{Zank14}
{Zank} G.~P.,  {le Roux} J.~A.,  {Webb} G.~M.,  {Dosch} A.,   {Khabarova} O.,
  2014, \mn@doi [\apj] {10.1088/0004-637X/797/1/28}, \href
  {http://adsabs.harvard.edu/abs/2014ApJ...797...28Z} {797, 28}

\bibitem[\protect\citeauthoryear{{Zank} et~al.,}{{Zank} et~al.}{2015}]{Zank15}
{Zank} G.~P.,  et~al., 2015, \mn@doi [\apj] {10.1088/0004-637X/814/2/137},
  \href {http://adsabs.harvard.edu/abs/2015ApJ...814..137Z} {814, 137}

\bibitem[\protect\citeauthoryear{{le Roux}, {Zank}, {Webb}  \& {Khabarova}}{{le
  Roux} et~al.}{2016}]{leroux16}
{le Roux} J.~A.,  {Zank} G.~P.,  {Webb} G.~M.,   {Khabarova} O.~V.,  2016,
  \mn@doi [\apj] {10.3847/0004-637X/827/1/47}, \href
  {http://adsabs.harvard.edu/abs/2016ApJ...827...47L} {827, 47}

\bibitem[\protect\citeauthoryear{{van Marle}, {Casse}  \& {Marcowith}}{{van
  Marle} et~al.}{2018}]{vanMarle18}
{van Marle} A.~J.,  {Casse} F.,   {Marcowith} A.,  2018, \mn@doi [\mnras]
  {10.1093/mnras/stx2509}, \href
  {http://adsabs.harvard.edu/abs/2018MNRAS.473.3394V} {473, 3394}

\makeatother
\end{thebibliography}
\bsp	
\label{lastpage}
\end{document}